\documentclass[fleqn,usenatbib]{mnras}

\usepackage[T1]{fontenc}

\DeclareRobustCommand{\VAN}[3]{#2}
\let\VANthebibliography\thebibliography
\def\thebibliography{\DeclareRobustCommand{\VAN}[3]{##3}\VANthebibliography}

\usepackage{graphicx}	
\usepackage{amsmath}	
\usepackage{amssymb}	
\usepackage{newtxtext,newtxmath}

\title[Vertical structure of debris disks]{The vertical structure of debris disks and the impact of gas.}

\author[J. Olofsson et al.]{
Johan Olofsson,$^{1,2,3}$\thanks{E-mail: johan.olofsson@uv.cl}
Philippe Th\'ebault,$^{4}$
Quentin Kral,$^{4}$
Amelia Bayo,$^{1,2}$
Anthony Boccaletti,$^{4}$\newauthor
Nicol\'as Godoy,$^{1,2}$
Thomas Henning,$^{4}$
Rob G. van Holstein,$^{5}$
Karina Mauc\'o,$^{1,2}$
Julien Milli,$^{6}$\newauthor
Mat\'ias Montesinos,$^{7,2}$
Hanno Rein,$^{8,9}$
Antranik A. Sefilian$^{2,10,11}$
\\
$^{1}$Instituto de F\'isica y Astronom\'ia, Facultad de Ciencias, Universidad de Valpara\'iso, Av. Gran Breta\~na 1111, Playa Ancha, Valpara\'iso, Chile\\
$^{2}$N\'ucleo Milenio de Formaci\'on Planetaria (NPF), Chile\\
$^{3}$Max Planck Institut f\"ur Astronomie, K\"onigstuhl 17, 69117 Heidelberg, Germany\\
$^{4}$LESIA-Observatoire de Paris, UPMC Univ. Paris 06, Univ. Paris-Diderot, France\\
$^{5}$European Southern Observatory, Alonso de Cordova 3107, Vitacura, Casilla 19001, Santiago, Chile\\
$^{6}$Univ. Grenoble Alpes, CNRS, IPAG, F-38000 Grenoble, France\\
$^{7}$Escuela de Ciencias, Universidad Vi\~na del Mar, Vi\~na del Mar, Chile.\\
$^{8}$Department of Physical and Environmental Sciences, University of Toronto at Scarborough, Toronto, Ontario M1C 1A4, Canada\\
$^{9}$Department of Astronomy and Astrophysics, University of Toronto, Toronto, Ontario, M5S 3H4, Canada\\
$^{10}$Department of Applied Mathematics and Theoretical Physics, University of Cambridge, Wilberforce Road, Cambridge CB3 0WA, UK\\
$^{11}$Center for Advanced Mathematical Sciences, American University of Beirut, PO Box 11-0236, Riad El-Solh, Beirut 11097 2020, Lebanon
}

\date{Accepted XXX. Received YYY; in original form ZZZ}

\pubyear{2021}

\begin{document}
\label{firstpage}
\pagerange{\pageref{firstpage}--\pageref{lastpage}}
\maketitle

\begin{abstract}
The vertical structure of debris disks provides clues about their dynamical evolution and the collision rate of the unseen planetesimals. Thanks to the ever-increasing angular resolution of contemporary instruments and facilities, we are beginning to constrain the scale height of a handful of debris disks, either at near-infrared or millimeter wavelengths. Nonetheless, this is often done for individual targets only. We present here the geometric modeling of eight disks close to edge-on, all observed with the same instrument (SPHERE) and using the same mode (dual-beam polarimetric imaging). Motivated by the presence of CO gas in two out of the eight disks, we then investigate the impact that gas can have on the scale height by performing N-body simulations including gas drag and collisions. We show that gas can quickly alter the dynamics of particles (both in the radial and vertical directions), otherwise governed by gravity and radiation pressure. We find that, in the presence of gas, particles smaller than a few tens of microns can efficiently settle toward the midplane at the same time as they migrate outward beyond the birth ring. For second generation gas ($M_\mathrm{gas} \leq 0.1$\,$M_\oplus$), the vertical settling should be best observed in scattered light images compared to observations at millimeter wavelengths. But if the gas has a primordial origin ($M_\mathrm{gas} \geq 1$\,$M_\oplus$), the disk will appear very flat both at near-infrared and sub-mm wavelengths. Finally, far beyond the birth ring, our results suggest that the surface brightness profile can be as shallow as $\sim -2.25$.
\end{abstract}




\section{Introduction}

The vertical scale height of circumstellar disks varies significantly as the central object evolves. Young, gas-rich disks usually display a flared vertical profile, but with considerable variation depending on the presence or lack of gaps (e.g., \citealp{Avenhaus2018}), with a non negligible scale height, which is governed by the temperature profile of the gas. At later stages, once most of the gas has been dissipated, debris disks appear much thinner than their progenitors, but are not as razor thin as the rings around planets of the solar system. For instance, \citet{Thebault2009} studied the vertical structure of debris disks, and concluded that disks should naturally have a non-zero vertical scale height due to radiation pressure and collisions, and that the scale height should depend on the size of the particles.

In the past years, we have witnessed a significant improvement on the angular resolution provided by several instruments or facilities. Extreme adaptive optics systems routinely provide exquisite observations at near-infrared (IR) wavelengths, while long baseline interferometry helps us achieve comparable resolution at millimeter (mm) wavelengths. In this paper, we first focus on what can currently be achieved using scattered, polarized light observations, with the aim of better characterizing the vertical structure of debris disks (Section\,\ref{sec:obs}).

Another remarkable recent result is the paradigm change regarding the absence of gas in debris disks, which has been severely challenged over the past years. This change of paradigm on the presence of gas in debris disks happened mostly thanks to sub-mm observations of CO lines done with APEX and ALMA (\citealp{Hughes2008}, \citealp{Kospal2013}, \citealp{Moor2015}, \citealp{Matra2019}, \citealp{Kral2020}, and \citealp{Schneiderman2021}  among others). The origin of the gas can in most cases be explained as second generation gas, even for the most massive gas disks that were initially thought to be of primordial origin (\citealp{Kral2017,Kral2019}). Not only CO (or isotopes of CO) have been detected, but also \ion{O}{I} (\citealp{Riviere2012}, \citealp{Brandeker2016}, \citealp{Kral2016}), \ion{C}{I} (\citealp{Cataldi2018,Cataldi2020}, \citealp{Kral2019}), and \ion{C}{II} (\citealp{Cataldi2014}), at far-infrared or sub-mm wavelengths. Nonetheless, other molecules such as HCN or HCO$^+$ seem to be largely absent in CO-rich debris disks (\citealp{Matra2018}, \citealp{Smirnov2021}, \citealp{Klusmeyer2021} ). More stochastic events such as falling evaporating bodies (\citealp{Beust2001}) can be detected using optical spectroscopic observations, tracing other lines (in absorption) such as \ion{Ca}{II} or \ion{Na}{I}, if the gas is in the line of sight of the star (e.g., \citealp{Kiefer2014}, \citealp{Rebollido2018}, \citealp{Iglesias2018,Iglesias2019}, \citealp{Rebollido2020}). Such detections inform us about the transport of volatiles in the vicinity of the star but do not trace the bulk reservoir of gas in the outer disk.

For younger, optically thick disks, it is well known that the gas drag force will cause millimeter-sized grains to migrate towards the central star (\citealp{Adachi1976}, \citealp{Weidenschilling1977}); as the gaseous disk is supported by its own pressure, it rotates at sub-Keplerian velocity. Dust particles that are not coupled with the gas, on the other hand, should rotate at Keplerian velocity and therefore feel a head-wind causing them to spiral inwards. For optically thin and gas-bearing debris disks the problem has also been investigated (\citealp{Takeuchi2001}, \citealp{Thebault2005}, \citealp{Krivov2009}). Because the effect of radiation pressure cannot be neglected in debris disks, especially because our general census is strongly biased towards intermediate-mass stars, the migration can be both inwards (for large grains) or outwards (for small grains). But most studies focused only on the radial effect that gas drag has on the dust grains. \citet{Krivov2009} only remarked that one should expect a decrease of the aspect ratio of the disk but without quantifying it. Indeed, the velocity of the gas should not have a vertical component, while the orbits of slightly inclined particles should cross the midplane twice per orbit. The dust particles should therefore experience a non negligible vertical component of the head-wind. In Sections\,\ref{sec:simuls} and \ref{sec:model}, we present how we perform numerical simulations and the results we obtain, trying to quantify the effect of gas drag on the vertical (and radial) structure of debris disks. In Section\,\ref{sec:discussion} we aim at putting the results of our simulations into a broader context before summarizing our findings in Section\,\ref{sec:conclusion}.

\section{Observational constraints on the vertical thickness}\label{sec:obs}

We first need to define a sample of debris disks for which we may be able to resolve their vertical scale height. Only a few disks have been (marginally) spatially resolved vertically using ALMA (e.g., AU\,Mic, \citealp{Daley2019} or HR\,4796, \citealp{Kennedy2018}), while several disks have been observed at near-IR wavelengths using second generation instruments such as the ``Spectro-Polarimetric High Contrast Exoplanet REsearch'' instrument (SPHERE, \citealp{Beuzit2019}). 
To constrain the vertical scale height, highly inclined (with respect to the line of sight) debris disks are best suited. We therefore focused our sample on bright disks with inclinations larger than $\sim 70^{\circ}$ that have been observed at high angular resolution. Since the angular differential imaging (ADI) technique can bias the determination of the width of the disk (\citealp{Milli2012}), we rather searched for polarimetric observations (dual-beam polarimetric imaging, DPI) obtained with SPHERE.

Our sample consists of eight disks; HD\,32297 (\citealp{Bhowmik2019}), HD\,106906 (\citealp{Kalas2015}, \citealp{Lagrange2016}, \citealp{vanHolstein2021}), HD\,120326 (\citealp{Bonnefoy2017}), HD\,61005 (\citealp{Esposito2016}, \citealp{Olofsson2016}), HR\,4796 (\citealp{Perrin2015}, \citealp{Milli2017,Milli2019}, \citealp{Olofsson2019}, \citealp{Arriaga2020}, \citealp{Chen2020}), AU\,Mic (\citealp{Boccaletti2015,Boccaletti2018}), HD\,129590 (\citealp{Matthews2017}), and HD\,115600 (\citealp{Currie2015}, \citealp{Gibbs2019}). All of those debris disks, except HD\,120326, were also presented in \citet{Esposito2020}. The observations used in this paper were obtained using IRDIS in DPI mode (\citealp{Dohlen2008}, \citealp{vanHolstein2020}, \citealp{deBoer2020}), except HR\,4796 for which we used the ZIMPOL (\citealp{Schmid2018}) observations presented in \citet{Milli2019}. The distances $d_\star$ for all eight stars are reported in Table\,\ref{tab:ddit}, and all come from the Gaia Early Data Release 3 (\citealp{Gaia2016,Gaia2020}).

Other debris disks that were considered but not included in the sample are the disks around $\beta$\,Pictoris (\citealp{Millar-Blanchar2015}, \citealp{vanHolstein2021}), 49\,Cet (\citealp{Pawellek2019}), HIP\,73145 (HD\,131835, \citealp{Feldt2017}), HD\,15115 (\citealp{Engler2019b}), HD\,117214 (\citealp{Engler2019}), and GSC\,07396-00759 (\citealp{Adam2021}), the latter four being detected at rather low signal to noise ratio in the available DPI observations. For $\beta$\,Pictoris, the warped inner disk strongly biases the determination of the vertical scale height of the outer disk, rendering the modeling challenging. The disk around 49\,Cet is overall very extended and very faint (\citealp{Pawellek2019}) and was therefore not considered in this study.

\subsection{Data reduction}

For HR\,4796, we did not re-process the observations and used the final azimuthal Stokes $Q_\phi$ and $U_\phi$ images presented in \citet{Milli2019} and \citet{Olofsson2020}. To ensure systematic and consistent data treatment, for the other datasets, obtained with IRDIS, we reduced the data using the \texttt{IRDAP}\footnote{Available at \url{https://irdap.readthedocs.io/en/latest/}} package described in \citet[][version 1.3.1]{vanHolstein2020}. After performing cosmetic pre-preprocessing of the raw frames, the sets of images with perpendicular directions of linear polarization (obtained using different angles of the half-wave plate) were combined to derive the Stokes images $Q$ and $U$, in turn used to obtain the final $Q_\phi$ and $U_\phi$ images (see \citealp{deBoer2020}). The pixel scale of the observations is $12.26$ and $7.2$\,mas for IRDIS and ZIMPOL, respectively.

For the IRDIS observations, we also measured the standard deviation (and full width at half maximum, FWHM) of the stellar point spread function (PSF) on the photometric frames. We fit a normal profile in the horizontal and vertical directions, for both the left and right camera of IRDIS, for all stars. On average, for the seven targets observed with IRDIS, we found a standard deviation of $1.61$\,pixels (FWHM of $3.8$\,pixels), comparable to the standard deviation of the ZIMPOL observations of HR\,4796 (a $30$\,mas FWHM, with a pixelscale of $7.2$\,mas/pixel yields a standard deviation of  $30/7.2/2.355 = 1.77$\,pixels, \citealp{Milli2019}).

\subsection{Modeling approach}

To estimate the vertical scale height of the debris disks, we used the same modeling approach as in \citet{Olofsson2020}. To summarize briefly, the code\footnote{Available at \url{https://github.com/joolof/DDiT}} produces synthetic images which are compared to the $Q_\phi$ images. The synthetic images are convolved following the strategy described in \citet{Engler2018} to better simulate how observations are obtained in polarimetric mode. The code computes the $Q_\phi$ image directly but when observing with SPHERE, we obtain linear polarization images along different directions ($Q^+$, $Q^-$, $U^+$, and $U^-$). Those frames are used to construct the $Q$ and $U$ images, and have been convolved by the instrumental PSF. The $Q$ and $U$ images are then used to derive the $Q_\phi$ and $U_\phi$ frames. To mimic this order of operation, the $Q_\phi$ image from the code is first transformed to synthetic $Q$ and $U$ images, which are then convolved with a 2D normal distribution with a standard deviation of $2$\,pixels (comparable with the observations). Afterwards, these convolved $Q$ and $U$ images are recombined into a $Q_\phi$ image. The observed $U_\phi$ images are used as a proxy for the uncertainties (though they can contain some astrophysical signal in some cases, \citealp{Heikamp2019}), which are estimated as the standard deviation in concentric annuli, with a width of two pixels. The only difference compared to the strategy presented in \citet{Olofsson2020} concerns the number density distribution $n_\mathrm{dens} \propto R_\mathrm{dens}(r) \times Z_\mathrm{dens}(r,z)$, more specifically, the shape of the vertical profile. Instead of using a normal distribution of the form
\begin{equation}
  Z_\mathrm{dens}(r, z) = \mathrm{exp}\left[-\frac{z^2}{2(\mathrm{tan}(\psi) r)^2} \right],
\end{equation}
where $\psi = \arctan(h/r)$ is the opening angle, $h$ the scale eight of the disk, $r$ the distance in the midplane, and $z$ the altitude, we now use an exponential fall-off of the form 
\begin{equation}\label{eqn:falloff}
  Z_\mathrm{dens}(r, z) = \mathrm{exp}\left[-\left(\frac{|z|}{\mathrm{tan}(\psi) r}\right)^{\gamma} \right],
\end{equation}
similarly to \citet{Artymowicz1989} and \citet{Augereau1999}. This functional form offers more control on the shape of the vertical profile, which is the prime focus of this work. If the vertical profile truly follows a gaussian profile, the modeling should yield a value of $\gamma = 2$ (then only the factor $2$ at the denominator will be missing compared to the previous implementation). We assume here that the disk has a constant opening angle and is not flared. The radial density distribution is fully described by three free parameters; the reference radius\footnote{As noted in \citet{Augereau1999}, this does not necessarily correspond to the radius of maximum density.} $a_0$, and two power-law slopes $\alpha_\mathrm{in}$ and $\alpha_\mathrm{out}$, as

\begin{equation}
    R_\mathrm{dens}(r) \propto \left[ \left(\frac{r}{a_0}\right)^{-2 \alpha_\mathrm{in}} + \left(\frac{r}{a_0}\right)^{-2 \alpha_\mathrm{out}}  \right]^{-1/2}.
\end{equation}

While some of the disks are known to be slightly eccentric, we limited the number of free parameters to seven; the reference radius $a_0$, its inclination $i$, position angle $\phi$, inner and outer slopes $\alpha_\mathrm{in}$ and $\alpha_\mathrm{out}$, opening angle $\psi$, and the factor $\gamma$ from Eqn.\,\ref{eqn:falloff}. We chose not to include the eccentricity and argument of pericenter as free parameters because a highly inclined disk is the least favorable configuration to constrain those two parameters: preliminary tests showed that to accommodate for small radial differences on both sides of the projected major-axis, the eccentricity can sometimes be artificially large (e.g., $e \sim 0.5$) if the argument of pericenter lies close to the line of sight (close to the projected semi-minor axis, either on the front or back sides). For the specific case of AU\,Mic, the east side is ``contaminated'' by the fast moving structures reported in \citet{Boccaletti2015,Boccaletti2018}, which renders the modeling of the disk challenging. For this reason, we elected to only compute the goodness of fit on the western side of this disk.

To find the best solution, we used the \texttt{Multinest} nested sampling algorithm (\citealp{Feroz2009,Feroz2019}), interfaced with \texttt{Python} using the \texttt{PyMultiNest} package (\citealp{Buchner2014}). When sampling the parameter space, for a given set of free parameters, we follow a two-step approach to evaluate the goodness of fit. A first model is computed, assuming an isotropic phase function, to only trace the dust density distribution, and not be affected by the phase function . For each pixel of the image, the scattering angle (the angle between the star, the dust grains, and the observer) is computed. The brightness profiles, as a function of the scattering angle, both for the observations and the dust density model, are estimated. For that set of parameters, the ``best'' phase function is equal to the profile from the observations divided by the profile from the dust density model (more details in \citealp{Olofsson2020}). This newly estimated phase function is then used to compute a second model, with the same input free parameters and the goodness of fit is estimated from the residuals. The main caveat of this approach is that the phase function is assumed to be the same as a function of the distance $r$ to the star (while for instance, \citealp{Stark2014} showed that this is not the case for the disk around HD\,181327). Since the spatial distribution of dust particles depends on their sizes, for a given scattering angle, the phase function that we derive is averaged over $r$, possibly leading to either over- or under-subtraction of the disk signal.  The sampler then picks another set of free parameters and the process is repeated until we obtain the best fit solution. 

For each disk, we define a range of uniform priors for each free parameter. The ranges of the priors were estimated from previously published values and are wide enough to ensure that the fitting algorithm does not reach a border (typically $\pm 0.25\arcsec$ for $a_0$, $\pm 10^{\circ}$ for $i$ and $\phi$). For $\alpha_\mathrm{in}$, $\alpha_\mathrm{out}$, $\psi$, and $\gamma$ the priors are set to $[1, 40]$, $[-25, -0.5]$, $[0.005, 0.1]$\,rad, and $[0.1, 10.0]$, respectively. The goodness of fit is estimated inside an elliptical mask (or half of it in the case of AU\,Mic), the size of which depends on initial values for $a_0$, $i$, and $\phi$ (taken from the literature). An inner numerical mask is also considered to avoid the most central regions of the $Q_\phi$ image, which can be noisy if the PSF subtraction is not perfect. 

\subsection{Best fit parameters}

\begin{figure*}
	\includegraphics[width=0.8\hsize]{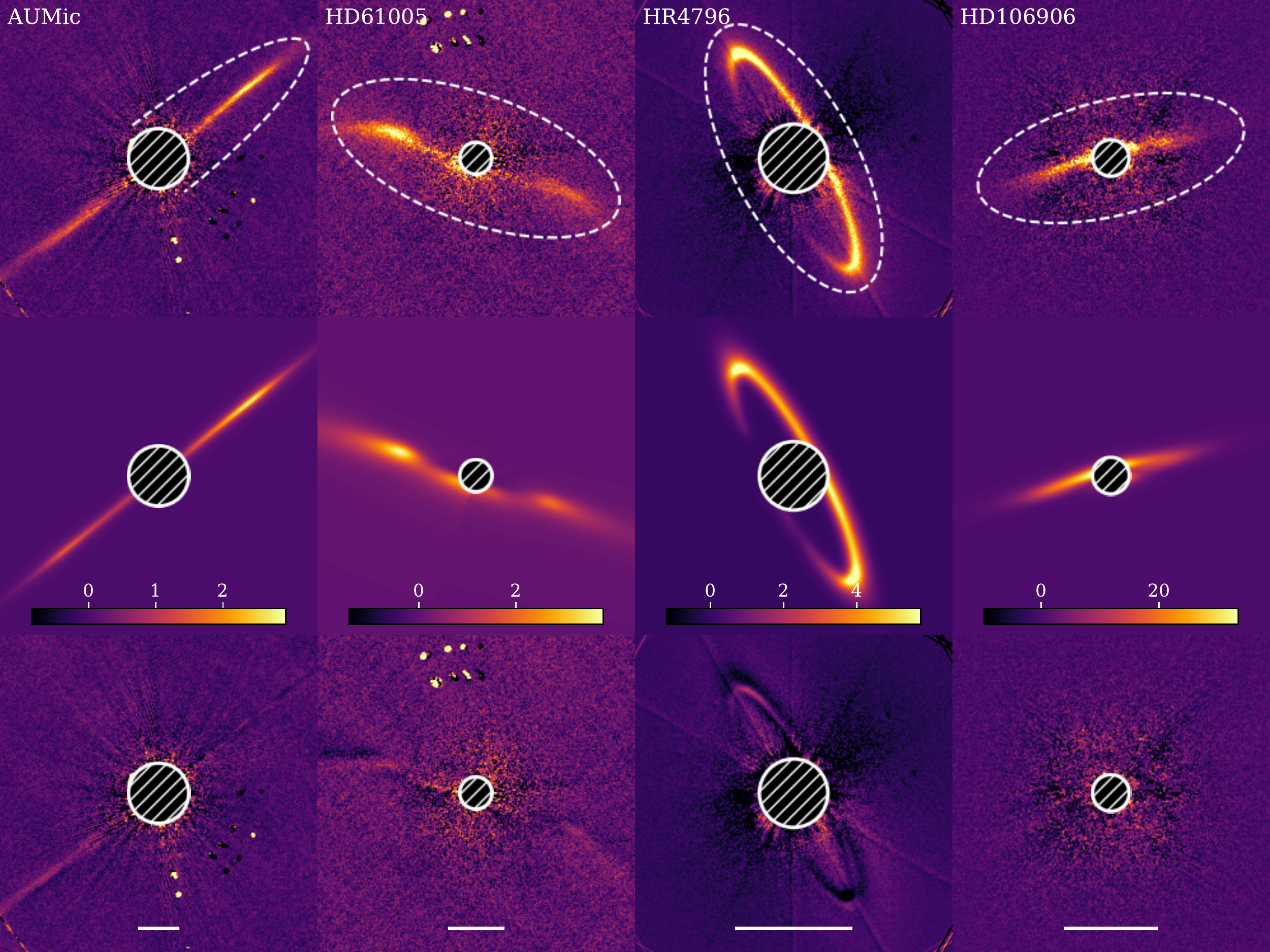}
	\includegraphics[width=0.8\hsize]{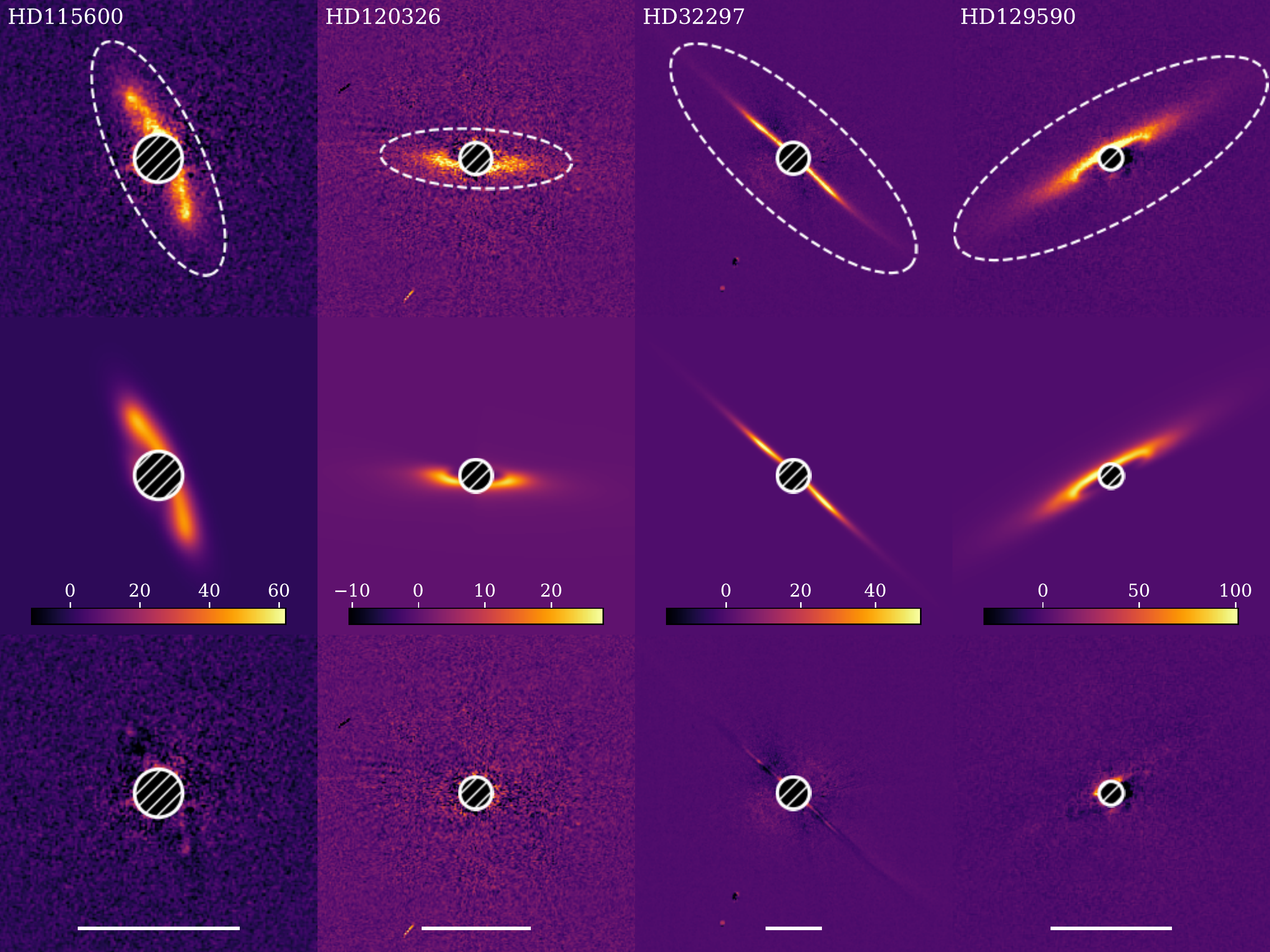}
    \caption{Top (and bottom) three panels, from top to bottom: observations ($Q_\phi$), best-fit model, and residuals for the geometric modeling. The stellar name is indicated in the upper panel. For each column, the scaling is linear and the same for the three panels (the colorbar is shown in the middle panel, the observations are not calibrated in flux). The elliptical mask within which the goodness of fit is estimated is shown in the top panel, but for AU\,Mic, only the western part of the disk is modeled. In the bottom panel, the horizontal bar represents $1\arcsec$.}
    \label{fig:all_first}
\end{figure*}

\begin{table*}
	\centering
    \caption{Results of the geometric modeling of the observations, ordered by increasing distance from earth. The columns show the stellar names, the distance, the reference radius $a_0$, the inclination $i$, position angle $\phi$, inner and outer slopes of the density distribution ($\alpha_\mathrm{in}$ and $\alpha_\mathrm{out}$, respectively), opening angle $\psi$, exponential fall-off $\gamma$, and the aspect ratio at the location of the reference radius. }
	\label{tab:ddit}
	\begin{tabular}{lccccccccc}
		\hline
        Star & $d_\star$ & $a_0$ & $i$ & $\phi$ & $\alpha_\mathrm{in}$ & $\alpha_\mathrm{out}$ & $\psi$ & $\gamma$ & $\mathrm{FWHM}/a_0$\\
             & [pc]      & [$\arcsec$] & [$^\circ$] & [$^\circ$] &  & & [rad] &  & \\
		\hline
AU\,Mic    & $9.714 \pm 0.002$  & $3.43 \pm 0.05$ & $89.4 \pm 0.2$ & $128.7 \pm 0.3$  & $7.7 \pm 0.4$  & $-4.2 \pm 0.1$ & $0.021 \pm 0.001$ & $0.87 \pm 0.04$ & $0.028 \pm 0.002$ \\
HD\,61005  & $36.45 \pm 0.02$  & $1.47 \pm 0.03$ & $82.6 \pm 0.1$ & $71.7 \pm 0.4$   & $6.6 \pm 0.4$  & $-1.4 \pm 0.2$ & $0.051 \pm 0.003$ & $1.99 \pm 0.07$ & $0.085 \pm 0.005$ \\
HR\,4796   & $70.77 \pm 0.24$  & $1.04 \pm 0.02$ & $77.7 \pm 0.2$ & $-151.6 \pm 0.3$ & $35.0 \pm 0.1$ & $-9.5 \pm 0.3$ & $0.041 \pm 0.002$ & $1.10 \pm 0.03$ & $0.059 \pm 0.002$ \\
HD\,106906 & $102.38 \pm 0.19$ & $0.85 \pm 0.03$ & $84.3 \pm 0.4$ & $-74.7 \pm 0.6$  & $1.7 \pm 0.2$  & $-3.7 \pm 0.2$ & $0.047 \pm 0.006$ & $2.27 \pm 0.57$ & $0.080 \pm 0.011$ \\
HD\,115600 & $109.04 \pm 0.25$ & $0.45 \pm 0.02$ & $76.6 \pm 0.3$ & $-156.0 \pm 0.2$ & $2.2 \pm 0.2$  & $-6.0 \pm 0.2$ & $0.132 \pm 0.004$ & $8.32 \pm 1.52$ & $0.254 \pm 0.008$ \\
HD\,120326 & $113.27 \pm 0.38$ & $0.31 \pm 0.03$ & $76.7 \pm 0.6$ & $86.4 \pm 0.7$   & $12.0 \pm 3.9$ & $-1.9 \pm 0.1$ & $0.005 \pm 0.002$ & $7.27 \pm 1.30$ & $0.010 \pm 0.004$ \\
HD\,32297  & $129.73 \pm 0.55$ & $0.93 \pm 0.02$ & $86.9 \pm 0.3$ & $-132.3 \pm 0.2$ & $2.9 \pm 0.2$  & $-3.0 \pm 0.3$ & $0.022 \pm 0.002$ & $9.64 \pm 0.44$ & $0.042 \pm 0.003$ \\
HD\,129590 & $136.32 \pm 0.44$ & $0.35 \pm 0.02$ & $82.0 \pm 0.2$ & $-60.6 \pm 0.3$  & $32.4 \pm 2.5$ & $-1.8 \pm 0.1$ & $0.001 \pm 0.001$ & $9.70 \pm 0.23$ & $0.002 \pm 0.001$ \\
		\hline
	\end{tabular}
\end{table*}

Figure\,\ref{fig:all_first} shows the observations, best-fit model, and the residuals from top to bottom, respectively, for all eight disks. The elliptical masks are displayed only on the observations, and for AU\,Mic it shows that the goodness of fit is only evaluated on the western side of the disk. The best fit parameters are summarized in Table\,\ref{tab:ddit}, and the last column shows the FWHM at $a_0$, divided by $a_0$, that is the aspect ratio of the best fit model. The FWHM is estimated, from the best-fit parameters as $2 a_0 \mathrm{tan}(\psi)[-\mathrm{ln}(0.5)]^{1/\gamma}$. We note that the uncertainties, derived from the posterior density distribution functions are very small, a topic that has been discussed in, among other, \citet{Mazoyer2020} and \citet{Olofsson2020}, and this most likely arises from under-estimating the uncertainties from the $U_\phi$ image. The main degeneracies in the modeling are related to the triplet ($a_0$, $\alpha_\mathrm{in}$, $\alpha_\mathrm{out}$) since these three parameters govern the distance at which the density is maximum. We also note that $\psi$ and $\gamma$ can be correlated with each other if the disks are spatially resolved in the vertical direction (see later). Finally, we also obtained the polarized phase functions for all eight disks, shown in Figure\,\ref{fig:pfunc}, but will not further discuss them, as the purpose of this study is not to constrain the properties of the small dust grains (the phase function of HD\,32297 will be further investigated in Olofsson et al. in prep).

Qualitatively, our best fit models are a very good match to the observations, except for some specific features. However, the swept-back wings of HD\,61005 and HD\,32297 are not well reproduced. For both disks, \citet{Hines2007}, \citet{Maness2009}, and \citet{Debes2009} suggested that such morphology could be the consequence of interactions between small dust particles and the local interstellar medium (further investigated in \citealp{Pastor2017} for HD\,61005). \citet{Esposito2016} proposed a model that could account for the wings of the disk around HD\,61005 (see also \citealp{Lee2016}), requiring an eccentric ring, with the pericenter close to the line of sight. Since we assumed circular disks, the best fit model cannot match those features. The same applies to HR\,4796; the disk is known to be eccentric, and in \citet{Olofsson2020} they obtained a better solution that accounts for most of the signal of the disk. The small positive residuals along the projected minor axis of the disk around HD\,129590 could be related to the determination of the phase function for small scattering angles, as mentioned earlier. 

The inner and outer slopes, $\alpha_\mathrm{in}$ and $\alpha_\mathrm{out}$, reported in Table\,\ref{tab:ddit} show that debris disks can display diverse morphologies. The disk around HD\,106906\footnote{Even though the best fit solution for $\alpha_\mathrm{in}$ is not at the border of the prior ($[1, 40]$), we checked that shallower values for this parameter did not yield a better $\chi^2$ value.} (along with HD\,32297 to some extent) seems to be rather broad, with a shallow inner slope, as opposed to the disks around HR\,4796 and HD\,129590 with very steep inner edges. It should be noted though that high inclination is not the ideal configuration to adequately constrain the sharpness of the inner edge due to projection effects. On the other hand, six of the eight disks display rather shallow outer slopes (the radial density $R_\mathrm{dens}$ being volumetric, the surface density follows $\alpha_\mathrm{out}+1$). The disk with the steepest distribution is HR\,4796 (see also \citealp{Milli2017}), while HD\,115600 lies in between with a rather shallow inner slope and steep outer distribution.

\begin{figure}
	\includegraphics[width=\columnwidth]{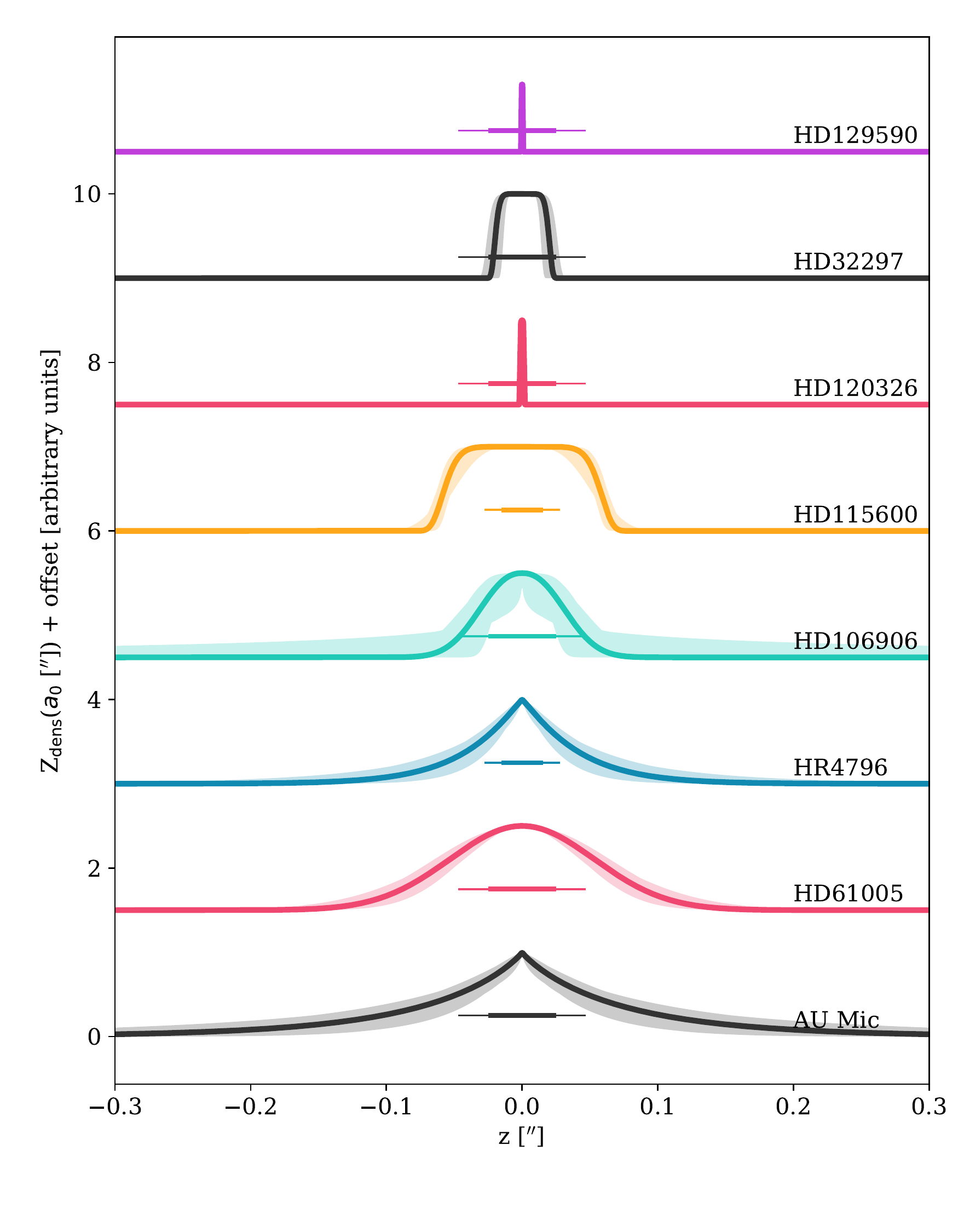}
    \caption{Comparison of the normalized vertical profiles derived from the best fit solutions for all eight disks (with offsets for clarity), at the reference radius $a_0$ (in units of arcseconds), from the closest to the farthest from earth (bottom to top, respectively). The shaded regions are representative of the uncertainties. For each profile, the horizontal bars show the width of one and two elements of resolution (thick and thin lines, respectively).}
    \label{fig:height_compare}
\end{figure}

Figure\,\ref{fig:height_compare} shows normalized vertical profiles for all eight disks (with offsets for clarity). The profiles are computed following Eqn.\,\ref{eqn:falloff}, using the best fit values for $a_0$ (in arcseconds), $\psi$, and $\gamma$. The shaded regions correspond to the minimum and maximum values of $1\,000$ profiles computed for different values of $\psi$ and $\gamma$ (drawn from normal distributions centered at the best fit values and with their uncertainties as the standard deviations). For each profile, the horizontal bars correspond to one and two elements of resolution (thick and thin lines, respectively) as estimated from the PSF ($3.8$\,pixels for the FWHM), scaled to the distances of the stars, accounting for the different pixel scales. The Figure shows the resolving power of the IRDIS and ZIMPOL instruments and suggests that the disks around AU\,Mic, HD\,61005, HR\,4796, and HD\,115600 are spatially resolved in the vertical direction, while the disks around HD\,106906, HD\,120326, HD\,32297, and HD\,129590 are not spatially resolved along the $z$ axis (or at best, marginally resolved for HD\,106906). We must note that HD\,129590, HD\,32297 and HD\,120326 are the three farthest disks in our sample. The width of the profiles for HD\,120326 and HD\,129590 on Fig.\,\ref{fig:height_compare} may appear very narrow, but the fitting process quickly converges to small and large values of $\psi$ and $\gamma$, respectively. The horizontal lines on the Figure denotes the size of one and two elements of resolution, but the width of the profiles (which are not convolved by a gaussian with a standard deviation of $2$\,pixels, as opposed to the model) is actually just smaller than the size of one pixel for both disks. Overall, the disks around those two stars are not resolved in the vertical direction but both seems to be very flat.

For the disks that are most likely resolved in the vertical direction, we find aspect ratios of FWHM$/a_0 = 0.028$, $0.085$, $0.059$, $0.254$, for AU\,Mic, HD\,61005, HR\,4796, and HD\,115600, respectively. Not only the aspect ratios are quite different, the shape of the profiles also show some diversity as displayed in Figure\,\ref{fig:height_compare}. The profiles for AU\,Mic and HR\,4796 show relatively broad wings, while the profile for HD\,61005 and HD\,115600 are either more gaussian-like ($\gamma \sim 2$) or appear ``boxy''. In this small sub-sample of four, only one system is known to host a planet, AU\,Mic (\citealp{Plavchan2020}), and all the central stars have different spectral types (from M1 to A0), which has implications on the strength of the radiation pressure (or stellar winds for low-mass stars). It is therefore challenging to draw broad conclusions on the aspect ratio of debris disks, but at a first glance, it seems that there is a significant diversity in the vertical scale height of the disks studied here (further discussed in Section\,\ref{sec:irscale}).

Some of the differences between the vertical structures of the different disks modeled in this Section may be due to the presence of gas. For the sample studied in this paper, stringent upper limits on the presence of CO gas were reported by \citet{Kral2020} for HD\,106906 and HR\,4796, by \citet{Lieman-Sifry2016} for HD\,115600, by \citet{Olofsson2016} for HD\,61005, and by \citet{Daley2019} for AU\,Mic. HD\,120326 has not been observed with ALMA and therefore there are no observational constraints regarding the presence of CO gas in this disk. The only two disks in our sample that harbor gas are HD\,32297 and HD\,129590. Both CO and \ion{C}{I} gas have been detected around the former (\citealp{Greaves2016}, \citealp{MacGregor2018}, \citealp{Moor2019}, and \citealp{Cataldi2020}) and CO gas has been detected around the latter (\citealp{Kral2020}). According to Table\,\ref{tab:ddit} the thinnest disks are around HD\,129590, HD\,120326, AU\,Mic, and HD\,32297. Even though three out of four of those disks are far from the earth, they seem to be very flat, and two out of four have gas (and the presence of gas in the third one is unknown). In the following, we perform simulations to investigate the effect that a gaseous disk can have on the dynamics of the dust particles.

\section{Gas drag simulations}\label{sec:simuls}

The interaction between gas and dust particles in debris disks has been studied in several papers in the past decades. In this study we present an extension of the works presented in \citet{Takeuchi2001}, \citet{Thebault2005}, and \citet{Krivov2009}, that takes into account the vertical dimension of the disk and also the crucial parameter that is the collisional lifetime of the dust. In its essence, our algorithm is similar to the DyCoSS code (\citealp{Thebault2012}), with the additional inclusion of the gas drag force. In brief, thousands of particles are released within a ``birth ring'' and we follow their evolution over time, accounting for gravity, radiation pressure and gas drag forces, taking into account their collisional lifetime.

\subsection{The gaseous disk}\label{sec:gas}

We aim at parameterizing the disk of gas with as few parameters as possible. The disk starts at a radius $r_0$, and follows a volumetric density profile proportional to $(r/r_0)^{-\alpha_\mathrm{gas}}$. To avoid boundary effects (see \citealp{Takeuchi2001}), we do not define an outer radius for the gaseous disk, but its mass density distribution is normalized to $M_\mathrm{gas}$ between the radii $r_0$ and $r_1$ (both $r_0$ and $r_1$ are chosen so that the gaseous disk is wider than the birth ring of planetesimals, see Section\,\ref{sec:dustdynamics}). The temperature profile of the gas also follows a power-law that goes as $T_0(r/r_0)^{-p}$ (i.e., without any dependency on $z$), where $T_0$ is the temperature at $r_0$. Following \citet{Armitage2015}, we assume the vertical density distribution to follow a normal distribution proportional to $\mathrm{exp}[ -z^2/(2h_\mathrm{gas}^2) ]$, where $h_\mathrm{gas}$ is the scale height of the gaseous disk, computed as
\begin{equation}\label{eqn:hgas}
    h_\mathrm{gas} = \frac{c_\mathrm{s}}{\Omega_\mathrm{k}} = \sqrt{\frac{k_\mathrm{b}T_0}{\mu m_\mathrm{H}}\left(\frac{r}{r_0}\right)^{-p}\frac{r^3}{G M_\star}} ,
\end{equation}
where $c_\mathrm{s}$ is the sound speed, $\Omega_\mathrm{{k}}$ the Keplerian angular velocity in the midplane, $k_\mathrm{b}$ the Boltzmann constant, $\mu$ the mean molecular weight, $m_\mathrm{H}$ the mass of the hydrogen atom, $G$ the gravitational constant, and $M_\star$ the stellar mass. Furthermore, we assume that the gaseous disk has zero viscosity and no sink, and therefore will not change throughout the simulations. We neglect the effect of self-gravity of the disk. For instance, \citet{Sefilian2021} showed that when accounting for the mass of the planetesimals themselves, self-gravity could induce additional precession and circularize the orbits of dust particles. If the gaseous disk is sufficiently massive, similar effects may also take place. We also neglect back-reaction between the gas and the dust phases, such as the photoelectric heating of the gas by the dust particles (\citealp{Lyra2013}), meaning that the gas-to-dust ratio should not be a critical parameter since gas and dust are not interacting with each other. Last but not least, stochastic events, such as coronal mass ejections, flares, stellar winds, which could be responsible for instance for the fast moving structures detected around AU\,Mic (\citealp{Boccaletti2015}), are also not accounted for, the impact of all the aforementioned processes being out of the scope of this paper

The velocity of the gas $v_\mathrm{gas}$, which is sub-Keplerian as it is supported by its own pressure, is estimated using Equation 28 of \citet{Armitage2015},
\begin{equation}\label{eqn:vgas}
  v_\mathrm{gas} = v_\mathrm{k} \left[1 - \frac{1}{4}\left(\frac{h_\mathrm{gas}}{r}\right)^2 \left(p + 2\zeta + 3 + p\frac{z^2}{h_\mathrm{gas}^2} \right)\right]
\end{equation}
where $v_\mathrm{k} = \Omega_\mathrm{k} r$ is the Keplerian velocity, $\zeta$ is the exponent of the surface density distribution of the gas. Since we assume that the vertical distribution of the gas follows a Gaussian profile, and that the gaseous disk is not flared, we replaced $\zeta$ by $\alpha_\mathrm{gas} - 1$ (both $\zeta$ and $\alpha_\mathrm{gas}$ being positive in our convention). We note that the velocity of the gas only has an azimuthal component, the components in the radial or vertical directions are null. To be able to compare the efficiency of gas drag to the one of stellar radiation pressure (see next subsection), we define $\eta$ such that 
\begin{equation}\label{eqn:vgas2}
  v_\mathrm{gas} = v_\mathrm{k} \sqrt{1-\eta}.
\end{equation}
We also compute the thermal velocity of the gas following \citet{Takeuchi2001}
\begin{equation}
  v_\mathrm{therm} = \frac{4}{3} \sqrt{\frac{8k_\mathrm{b} T_0}{\pi \mu m_\mathrm{H}}\left(\frac{r}{r_0}\right)^{-p}}
\end{equation}

Overall, the free parameters that control the distribution of the gas around the star are the following: $r_0$, $r_1$, $T_0$, $M_\mathrm{gas}$, $M_\star$, $\mu$, $\alpha_\mathrm{gas}$, and $p$.

\subsection{Dust dynamics}\label{sec:dustdynamics}

We consider that dust particles are steadily collisionally produced from a birth ring of larger parent bodies. For the sake of simplicity, we do not use the two-power law approach used for the geometrical modeling but instead consider that the semi-major axis $a$ of the particles is randomly drawn from a power-law distribution with a slope $-\alpha_\mathrm{dust}$, between $a_\mathrm{min}$ and $a_\mathrm{max}$. Initially, the particles are on circular orbit, and for each particle, its inclination is drawn from a normal distribution, centered at $0$\,radians with a standard deviation of $0.05$\,radians (\citealp{Hughes2018}).

The particle size distribution is assumed to follow an idealized collisional equilibrium law in $\mathrm{d}n \propto s^{-3.5} \mathrm{d}s$ between $s_\mathrm{min}$ and $s_\mathrm{max}$. To avoid drawing too many small grains compared to large grains, the slope of the size distribution is in fact $-1.4$ and a correction factor is applied a posteriori to account for the difference with a slope in $-3.5$ when computing either the optical depth or synthetic images (see later). For each particle, we compute the ratio $\beta$ between the radiation pressure and gravitational forces, similarly to \citet{Krivov2009}
\begin{equation}
  \beta = 0.5738 Q_\mathrm{pr} \left( \frac{1 \mathrm{g}.\mathrm{cm}^{-3}}{\rho_\mathrm{dust}} \right) \left(\frac{1 \mu\mathrm{m}}{s} \right) \frac{L_\star}{M_\star},
\end{equation}
where $Q_\mathrm{pr}$ is the radiation pressure efficiency (set to unity as in \citealp{Krivov2009}), $\rho_\mathrm{dust}$ the density of the dust particles (set to $3.3$\,g.cm$^{-3}$), $L_\star$, and $M_\star$ the stellar luminosity and mass (in units of solar luminosity and mass). 

\begin{figure}
	\includegraphics[width=\columnwidth]{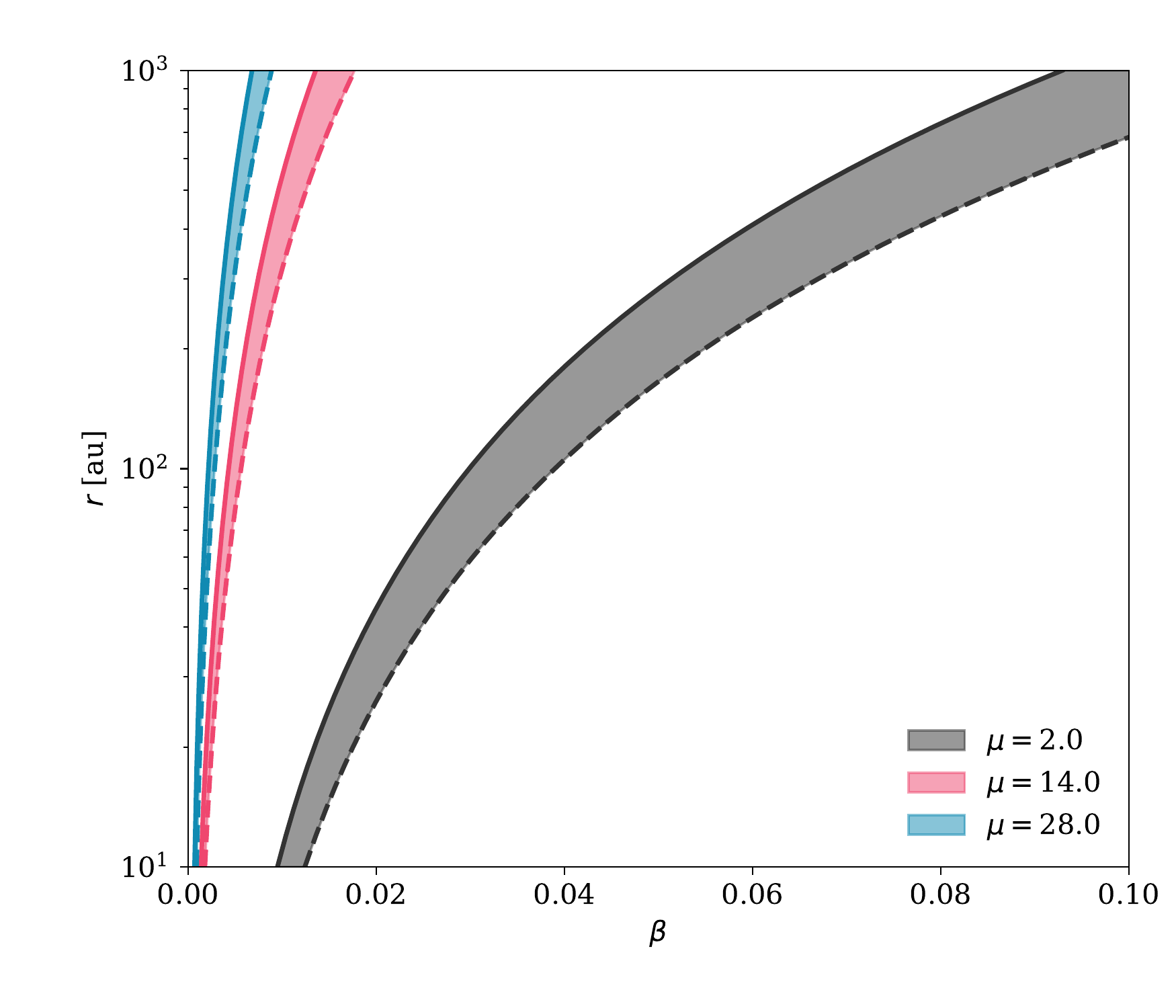}
    \caption{Radial location of the limit between inward and outward migration as a function of the radial distance and grain size (parametrized with $\beta$). Each colored region corresponds to different values of $\mu$ ($2$, $14$, $28$ in gray, red, and blue, respectively). For each region, the solid line corresponds to volumetric surface density following a radial slope of index $\alpha_\mathrm{gas} = -2.5$, while the dashed line corresponds to $\alpha_\mathrm{gas} = -3.5$. Particles on the right side of a line will migrate outward, particles to the left side of a line will migrate inward.}
    \label{fig:stability}
\end{figure}

Only three forces are governing the motion of the particles, stellar gravity, which is reduced by a factor $(1-\beta)$ due to radiation pressure, and the gas drag force. The latter is computed following \citet{Takeuchi2001} as
\begin{equation}
F_\mathrm{gas} = - \pi \rho_\mathrm{gas} s^2 \sqrt{v_\mathrm{therm}^2 + \Delta v^2} \Delta v,
\end{equation}
where $\rho_\mathrm{gas}$ is the local density of the gas disk (as computed in Section\,\ref{sec:gas}) and $\Delta v = v_\mathrm{dust} - v_\mathrm{gas}$ is the difference between the velocity of the dust particle and the velocity of the gas. For a particle of size $s$, if $\beta$ is larger than $\eta$ (from Equation\,\ref{eqn:vgas}), then the grain will migrate outwards, and inwards if $\beta < \eta$. Figure\,\ref{fig:stability} shows the regions of inward or outward migration, for different values of $\mu$ (color-coded) and $\alpha_\mathrm{gas}$ (solid and dashed lines for $-2.5$ and $-3.5$, respectively). For a given gaseous disk model ($\mu$ and $\alpha_\mathrm{gas}$) a particle that lies at $(\beta, r)$ will migrate outward (inward) if it is located to the right (left) of the lines. The Figure shows that a wider range of grains sizes will migrate outward for larger values of the molecular weight $\mu$ , which should likely correspond to disks containing non-primordial second generation gas (\citealp{Kral2017}).

To follow the motion of particles in time, we used the C implementation of \texttt{rebound}\footnote{Available at \url{https://github.com/hannorein/rebound}} (\citealp{rebound}). All dust grains are considered as test particles, that do not interact with each other. Both the radiation pressure and the gas drag forces are implemented in a separate function that the code will call at each iteration. We used the leapfrog integrator (making sure not to round up or down the integration time), with a constant time step $\Delta t$ set to $10$\,years (see App.\,\ref{sec:app_dt} for a comparison with a shorter timestep). Every $100$\,years, the 3D positions and velocities are saved in a file. 

\subsection{Collisional lifetimes}

The procedure described in the previous subsection lacks a crucial ingredient, namely the collisional lifetime of the produced dust. Without it, there can be an unrealistic and almost limit-less accumulation of particles in certain regions of the disk, in particular those where the grains' outbound and inbound motions almost balance each other. The final result from such a simulation would then heavily depend on when we (arbitrarily) stop it. We thus take into account collisional lifetimes, by implementing a procedure that is similar to that developed for the DyCoSS code by \citet{Thebault2012}. This procedure is iterative and makes the simplifying assumption that collisions are fully destructive. It can be summarized as follows:

We first run a collisionless simulation, referred to as ``run 0'', with $200\,000$ particles, for a total duration of $1\,000\,000$\,years, saving the positions of all particles at regularly spaced time steps. From the cumulated $3$D positions, we then estimate a $2$D optical depth map (radial and vertical) , which we normalize so that the 10\% densest regions have a median vertical optical depth of $\tau = 5 \times 10^{-3}$, typical for the bright debris disks considered here. The normalized optical depth map (equivalent to a dust mass distribution) is then used to estimate collisional destruction probabilities, as a function of disc location, in the next run (``run 1''). At each timestep $\Delta t$, this destruction probability is estimated with the simplified formula:
\begin{equation}
  f_\mathrm{coll} = \frac{\pi \Delta t \tau(r, \psi_\mathrm{c})}{t_\mathrm{orbit}},
\end{equation}
where $\psi_\mathrm{c} = |\mathrm{arctan}(z/r)|$.

As in ``run 0'', we store the particles' positions at regularly spaced time intervals, with the important difference that we now only take into account particles that have not been removed by collisions yet. The simulation then naturally stops once all particles have been eliminated, either by collisions or because they have spiraled inward of $a_\mathrm{min}$\footnote{The inward migration of large grains is out of the scope of this paper.} or migrated outward to reach a pericenter greater than $1\,000$ au. From the cumulated positions, we derive a new 2D optical depth map, which we normalize the same way as for ``run 0''. This normalized map is then used to estimate collisional lifetimes in the next run (``run 2''), and the procedure is repeated 20 times in total and we check that the sum over the whole optical depth map of the last run does not differ by more than $0.1$\% compared to the previous one. Examples of such optical depth maps are shown in Fig.\,\ref{fig:tau_map}.

\subsection{Deriving synthetic images}

Once a run of a simulation is finished we can create synthetic images either in scattered light or at millimeter wavelengths, probing the thermal emission of the disk. Those images will be used to perform further diagnostics, such as constraining the vertical scale height of the disk or computing the surface brightness profiles. 

For scattered light synthetic images, assuming a certain pixel size and distance of the star, we take each position of each particle (saved in units of au) that has been saved during the run and find the pixel corresponding to this position, accounting for projection and rotation effects due to the inclination and position angle. We then increment the value of that pixel by the geometric cross section of the dust grain and divide by $1/\pi r^2$ to account for illumination effects. For the sake of simplicity, we consider isotropic scattering for all dust particles when computing the images, as the phase function will not impact the vertical scale height nor the surface brightness of face-on disks.

For millimetric synthetic images, we first need to compute the temperature of the dust particles, as a function of their sizes and distance to the star. To this end, we first compute the absorption coefficients $Q_\mathrm{abs}(\lambda, s)$ of the dust particles between $s_\mathrm{min}$ and $s_\mathrm{max}$ over a spectral range covering from optical to mm wavelengths. We use the optical constants from \citet[][astronomical silicates]{Draine2003} and the Mie theory to compute the absorption efficiencies. Assuming a stellar radius $R_\star$ and a stellar spectrum $F_\star$, the relationship between the distance to the star $r$ and the temperature of the dust particle $T_\mathrm{dust}$ can be computed as

\begin{equation}
  r(s, T_\mathrm{dust}) = \frac{R_\star}{2}\sqrt{\frac{\int_\lambda F_\star(\lambda) Q_\mathrm{abs}(\lambda, s)\mathrm{d}\lambda}{\int_\lambda \pi B_\lambda(T_\mathrm{dust}, \lambda) Q_\mathrm{abs}(\lambda,s)\mathrm{d}\lambda}},
\end{equation}
where $B_\lambda$ is the Planck function. For a given grain size and distance to the star, one can invert this relationship to derive the temperature of the particle. Once the pixel of the image corresponding to the position of the particle is found, its contribution is incremented by $4\pi s^2 Q_\mathrm{abs} \pi B_\lambda(T_\mathrm{dust})$ at the chosen wavelength (there should be an additional $1/4\pi d_\star^2$ factor, but since it is the same for all particles, we simply omit it). By default the wavelength of the simulation is set to $870$\,$\mu$m (corresponding to Band\,7 observations with ALMA).

\section{Gas modeling results}\label{sec:model}

To run the simulations, we first need to determine the stellar parameters, especially the stellar luminosity and mass. We used HD\,32297 as our reference star and run a set of simulations with the aim of capturing how different the outputs are for different parameters. For the stellar parameters, we used the ``Virtual Observatory SED Analyzer'' service (VOSA\footnote{\url{http://svo2.cab.inta-csic.es/theory/vosa/index.php}}, \citealp{Bayo2008}) to gather photometric measurements of HD\,32297, and derive the stellar parameters. The Gaia Early Data Release 3 (\citealp{Gaia2016,Gaia2020}) distance is $129.73 \pm 0.55$\,pc and we obtained the following parameters; extinction $A_\mathrm{v} = 0$, effective temperature $T_\mathrm{eff} = 8\,000 \pm 125$\,K, surface gravity $\mathrm{log}g = 5.0$, luminosity $L_\star = 7.61 \pm 0.13$\,$L_\odot$, and stellar mass $M_\star = 1.78$\,$M_\odot$. It should be noted that the stellar parameters in the end do not matter much, since the motion of the particles will only depend on $\beta$, and how it compares to $\eta$ at a given distance from the star (Fig.\,\ref{fig:stability}) 

\begin{table}
	\centering
    \caption{Summary of the twelve simulations analyzed, the one parameter that is changing is highlighted in bold font (simulation \#5 is the fiducial one).}
	\label{tab:models}
	\begin{tabular}{lcccc}
		\hline
        ID & $M_\mathrm{gas}$ & $\alpha_\mathrm{gas}$ & $\mu$ & $\tau$ \\
             & [$M_\oplus$] &  & &\\
		\hline
        \#1  & \textbf{0}    &  2.5 &  28 & $5 \times 10^{-3}$ \\
        \#2  & \textbf{10$^{-4}$} &  2.5 &  28 & $5 \times 10^{-3}$ \\
        \#3  & \textbf{10$^{-3}$} &  2.5 &  28 & $5 \times 10^{-3}$ \\
        \#4  & \textbf{0.01} &  2.5 &  28 & $5 \times 10^{-3}$ \\
        \#5  & 0.1  &  2.5 &  2 & $5 \times 10^{-3}$ \\
        \#6  & \textbf{0.2}  &  2.5 &  28 & $5 \times 10^{-3}$ \\
        \#7  & 0.1  &  \textbf{1.5} &  28 & $5 \times 10^{-3}$ \\
        \#8  & 0.1  &  \textbf{3.5} &  28 & $5 \times 10^{-3}$ \\
        \#9  & 0.1  &  2.5 & \textbf{2} & $5 \times 10^{-3}$ \\
        \#10 & 0.1  &  2.5 &  28 & \textbf{10$^{-2}$} \\
        \#11 & \textbf{1}  &  2.5 & 28 & $5 \times 10^{-3}$ \\
        \#12 & \textbf{10}  &  2.5 & \textbf{2} & $5 \times 10^{-3}$ \\
      \hline

	\end{tabular}
\end{table}

The nominal values for the parameters in the simulations are the following; for $a_\mathrm{min}$ and $a_\mathrm{max}$ (location of the birth ring) we took the values of $R_\mathrm{in}$ and $R_\mathrm{out}$ reported in \citet{MacGregor2018}, that is $78.5$ and $122$\,au, respectively, with a slope for the density distribution $\alpha_\mathrm{dust}$ in $3.5$. For the gaseous disk, we chose $r_0 = 70$\,au and $r_1 = 130$\,au, slightly smaller and larger values compared to the birth ring, to ensure that the dusty disk is well within the gaseous disk when the simulation starts (but, as a reminder, $r_1$ is only used for the gas mass normalization, and the gaseous disk extends beyond $r_1$). For the other parameters related to the gas disk, the default values are $\alpha=2.5$, $T_0 = 40$\,K at $r_0$, $p=0.5$ (similarly to \citealp{Krivov2009}, additional tests showed that varying $p$ between $0$ and $1$ has a negligible impact on the results), and $\mu = 28$ (the latter value being best suited for second generation gas, as shielded CO can dominate the total gas mass). Concerning the grain sizes, we set $s_\mathrm{min} = 1.5$\,$\mu$m which corresponds to $\beta = 0.496$ just below the critical value of $0.5$, while for $s_\mathrm{max}$ we chose $1$\,mm so that we sample a wide range of $\beta$ values (down to $\beta \sim 7\times 10^{-4}$). For each simulation, we initially release $200\,000$ particles with inclinations drawn from a normal distribution with a standard deviation of $0.05$\,radians. Finally, the default value to normalize the optical depth map is $5 \times 10^{-3}$ (but a larger dust mass is tested in simulation \#10 with $\tau = 10^{-2}$). We then run a set of twelve simulations in total, varying only one parameter at a time (except for simulation \#12 where we changed both $M_\mathrm{gas}$ and $\mu$), and the different simulations are summarized in Table\,\ref{tab:models}. Simulations \#11 and \#12, with larger gas masses (and a different value of $\mu$ for \#12) are meant to be more representative of primordial gaseous disks, while the other ten simulations would rather correspond to gaseous disks of secondary origin. For those last two simulations, the gas drage force is stronger, and we therefore used a shorter $\Delta t = 1$\,year instead of $10$\,years.

\begin{figure*}
	\includegraphics[width=\hsize]{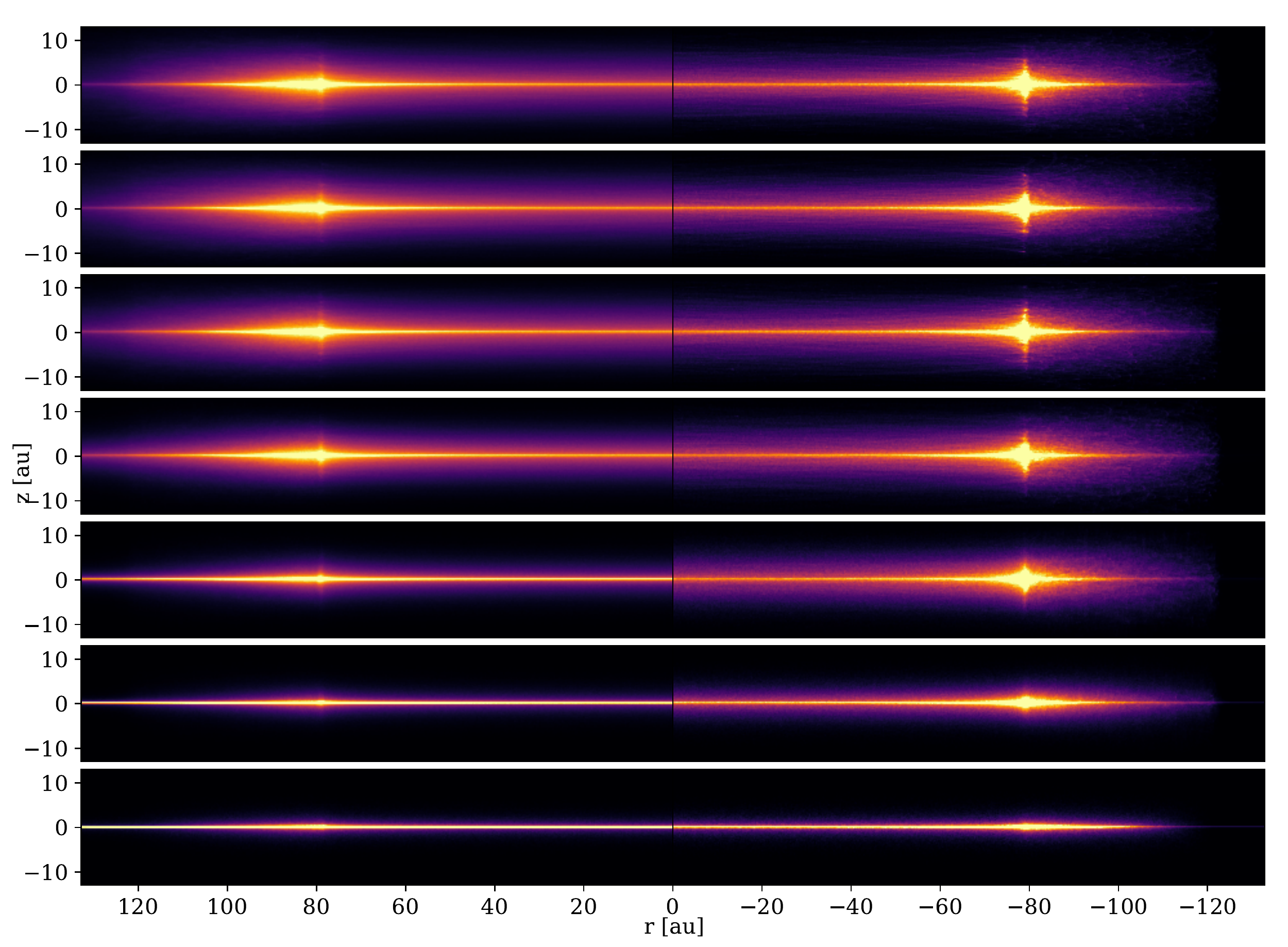}
    \caption{Synthetic images viewed edge-on for simulations \#1, \#2, \#3, \#4, \#5, \#11, and \#12 corresponding to gas masses of $M_\mathrm{gas} = 0$, $10^{-4}$, $10^{-3}$, $10^{-2}$, $0.1$, $1$, and $10$\,$M_\oplus$ (top to bottom respectively). The star is not included in the images. Each panel shows the scattered light image on the left, and the mm image on the right. The color scale is linear but for each sub-panel the maximum value is the $99$ percentile for that particular image. The pixel size is $1$\,mas.}
    \label{fig:edgeon}
\end{figure*}

Figure\,\ref{fig:edgeon} shows scattered light and mm synthetic images (left and right, respectively) for seven simulations with $M_\mathrm{gas} = 0$, $10^{-4}$, $10^{-3}$, $10^{-2}$, $0.1$, $1$, and $10$\,$M_\oplus$ (simulations \#1, \#2, \#3, \#4, \#5, \#11, and \#12), viewed edge-on, from top to bottom respectively. The pixel scale is set to $1$\,mas and the distance to $129.73$\,pc ($0.13$\,au per pixel). Immediately, the difference on the vertical scale height in scattered light is noticeable when increasing the gas mass in the system. Gas drag seems to have little to no effect on the mm images until we reach a mass of $1$\,$M_\oplus$, and the disk becomes thinner as the gas mass increases.

\subsection{Eccentricity and inclination damping}

\begin{figure*}
	\includegraphics[width=\hsize]{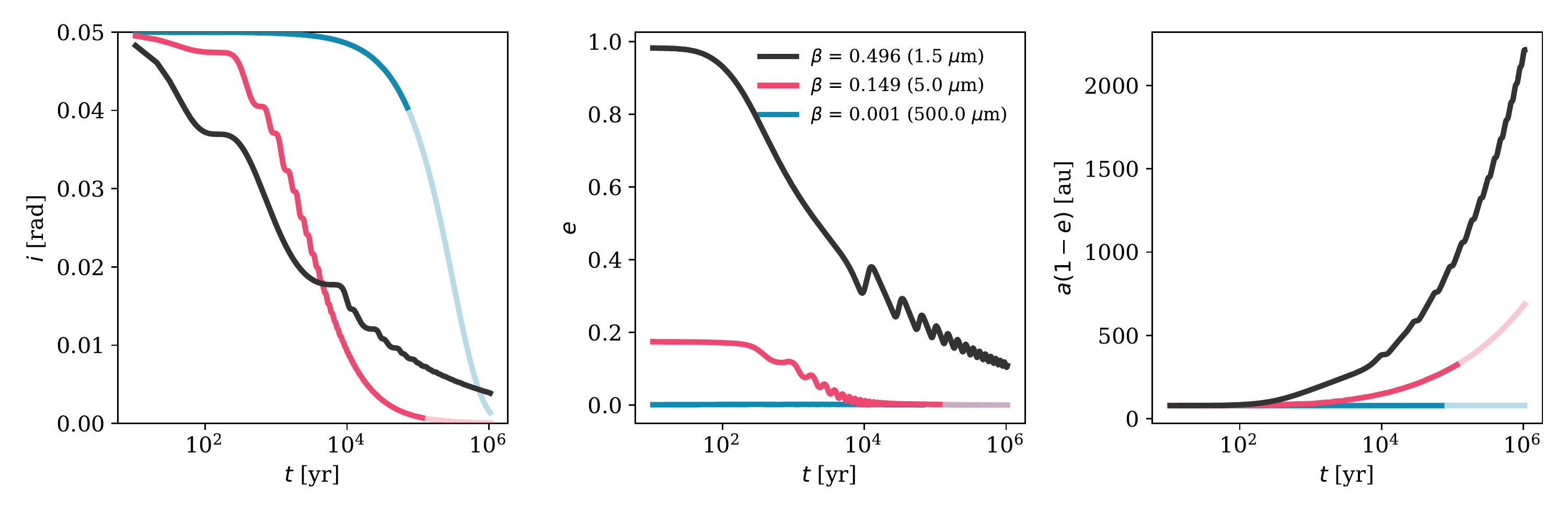}
    \caption{Evolution of the inclination $i$, eccentricity $e$, and distance of the pericenter ($a \times (1-e)$, left to right, respectively) as a function of time, for three different values of $\beta$, for $M_\mathrm{gas} = 0.1$\,$M_\oplus$ and $\alpha = 2.5$. If the particle is destroyed during the simulation the orbital parameters are showed with larger transparency.}
    \label{fig:three_sizes}
\end{figure*}

In the absence of gas, upon release, the eccentricity of a particle with a given $\beta$ should jump to $e = \beta / (1 - \beta)$ and remain unchanged thereafter (as is the case for the other orbital parameters). But accounting for the force exerted by the gaseous disk will introduce a time dependence on the orbital parameters. Figure\,\ref{fig:three_sizes} shows the evolution of $i$, $e$, and the location of pericenter $a(1-e)$ for three different values of $\beta$, for simulation \#5. For this Figure, we also checked whether the particle is eventually destroyed or not, and if this happens, the lines are shown with larger transparency (we remark though that the collisional lifetime does depend on the initial conditions when the particles are released). For all three sizes, both the inclinations and eccentricities of the particles decrease before reaching an asymptotic behavior that will ultimately reach zero. The orbits of the dust particles will therefore become more circular (if they can survive that long) due to the effect of gas drag (see Section\,\ref{sec:settling} for further discussion on the inclination). For the curves corresponding to $\mu$m-sized dust grains, in the left and center panels of Fig.\,\ref{fig:three_sizes}, oscillations can be seen. Those oscillations correspond to when a particle crosses the midplane of the disk (twice per orbit) as it is where the gas density is maximum. For the larger grains ($s = 0.5$\,mm), the eccentricity is negligible throughout the simulation, while the inclination steadily decreases (the $x$-axis being in log-scale) with time before reaching $0$.

The rightmost panel Fig.\,\ref{fig:three_sizes} shows that the location of the pericenter will drift to larger distances as a function of time and that this behavior is more pronounced for larger $\beta$ values, justifying our criteria for removing particles drifting too far away from the star.

\begin{figure}
	\includegraphics[width=\hsize]{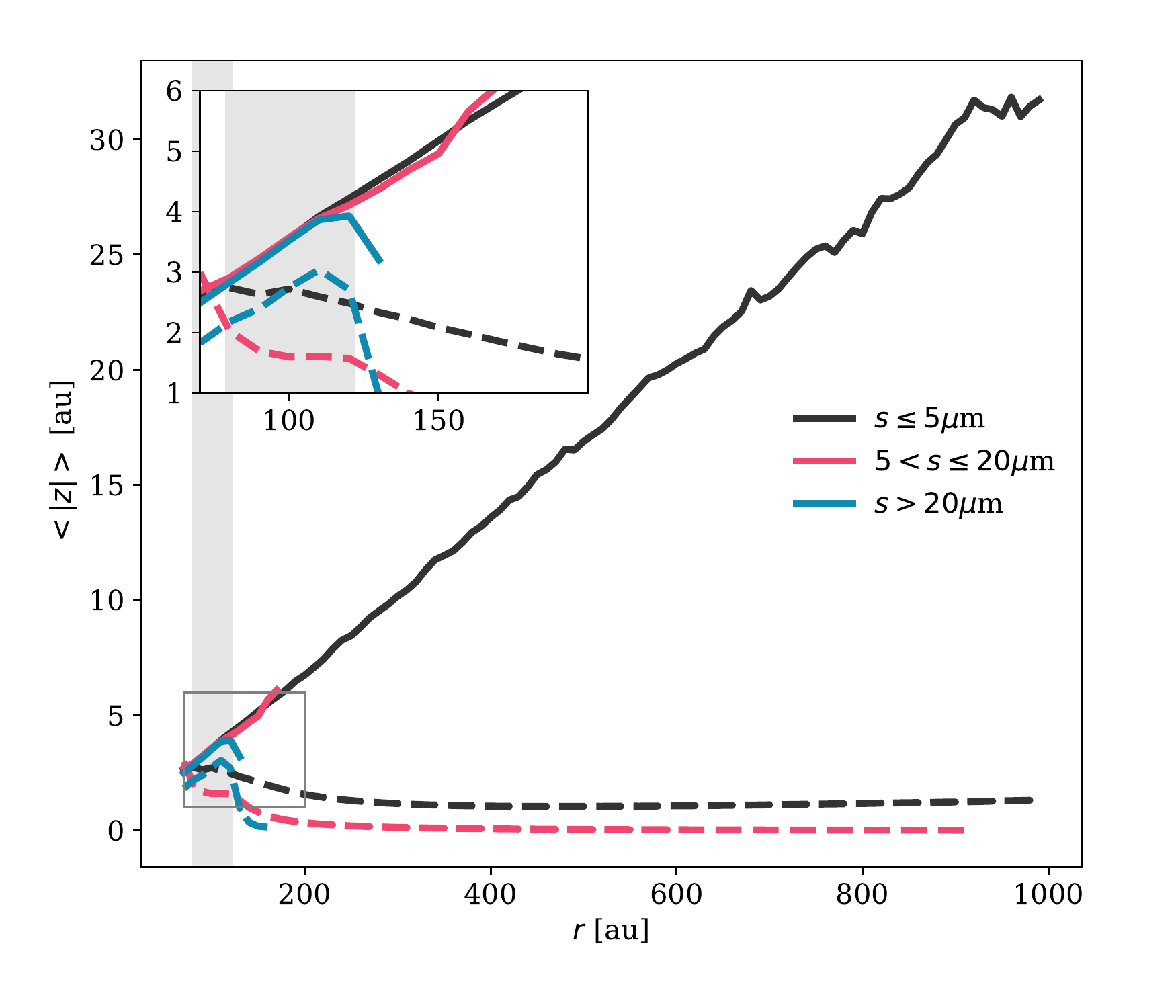}
    \caption{Mean absolute value for the altitude of particles as a function of the radial distance in the midplane, for several grain size bins for simulations \#1 and \#5, without gas and with $10^{-1}$\,$M_\oplus$ (solid and dashed lines, respectively). The inset shows a zoom on the inner regions. On both the main panel and the inset, the shaded area shows the location of the birth ring.}
    \label{fig:zcut}
\end{figure}

To better visualize how this translates when adding many more particles, for simulations \#1 and \#5, we compute the mean of the absolute value of the altitude $z$ as a function of the distance in the midplane ($\sqrt{x^2+y^2}$), for three different size intervals. Figure\,\ref{fig:zcut} shows the results for a simulation without gas (solid lines) and a simulation with a gas mass of $10^{-1}$\,$M_\oplus$ (dashed lines, simulation \#5). We divide particles into three size domains: $s\leq 5$\,$\mu$m, $5 \leq s \leq 20$\,$\mu$m, and $s > 20$\,$\mu$m. For the gas-free simulation, as expected, particles keep the orbital elements at the moment of their release, so that the mean absolute value of $z$ does not depend on their size (except that large grains do not venture far from the birth ring, because of their very low radiation pressure-induced eccentricity). When including gas in the simulation we see a departure from this behavior, and the vertical positions become dependent on particle sizes. Rather counter-intuitively, the mean $|z|$ value is smallest for the intermediate-size particles. While the smallest particles have a much smaller altitude compared to the simulation without gas, they are on average located at slightly higher $|z|$ compared to the intermediate-size grains. This can be better understood looking at Figure\,\ref{fig:three_sizes}: particles with intermediate value of $\beta$ (e.g., $\sim 0.2$) are set on orbits that are less eccentric than grains with larger $\beta$ values. Since the density of the gaseous disk decreases with stellar distance, this means that high-$\beta$ particles spend more time in lower gas density regions than intermediate-beta ones. Except for the initial $\sim 10^3$years, this effect dominates the overall effect of gas drag on the vertical distribution of particles, compensating for the $1/s$ dependence of the drag force. It explains why the inclination of $5$\,$\mu$m grains decreases faster than $1.5$\,$\mu$m one. As a consequence, and because these medium-$\beta$ grains survive long enough, their mean altitude is lower than that of particles closer to the blow-out size\footnote{For in-plane (i.e., horizontal) motions, this effect is mitigated by the fact that the initial high-$e$ of the smallest particles increases their relative velocity with gas streamlines. However, the gradual increase of the periastron of high-$\beta$ grains (which is much faster than that of medium-$\beta$ ones) makes that, in the long run, the orbits of $1.5$\,$\mu$m grains are less efficiently circularized than that of $5$\,$\mu$m ones.}. Overall, this exercise shows that the gas can have a significant impact on the orbits of the small and intermediate-sized dust particles, but the effect is less significant on particles larger than $20$\,$\mu$m (at least for this set of parameters).

\subsection{Vertical profiles and scale height}\label{sec:settling}

\subsubsection{Vertical profile: normal distribution or exponential fall-off?}

\begin{figure}
	\includegraphics[width=\hsize]{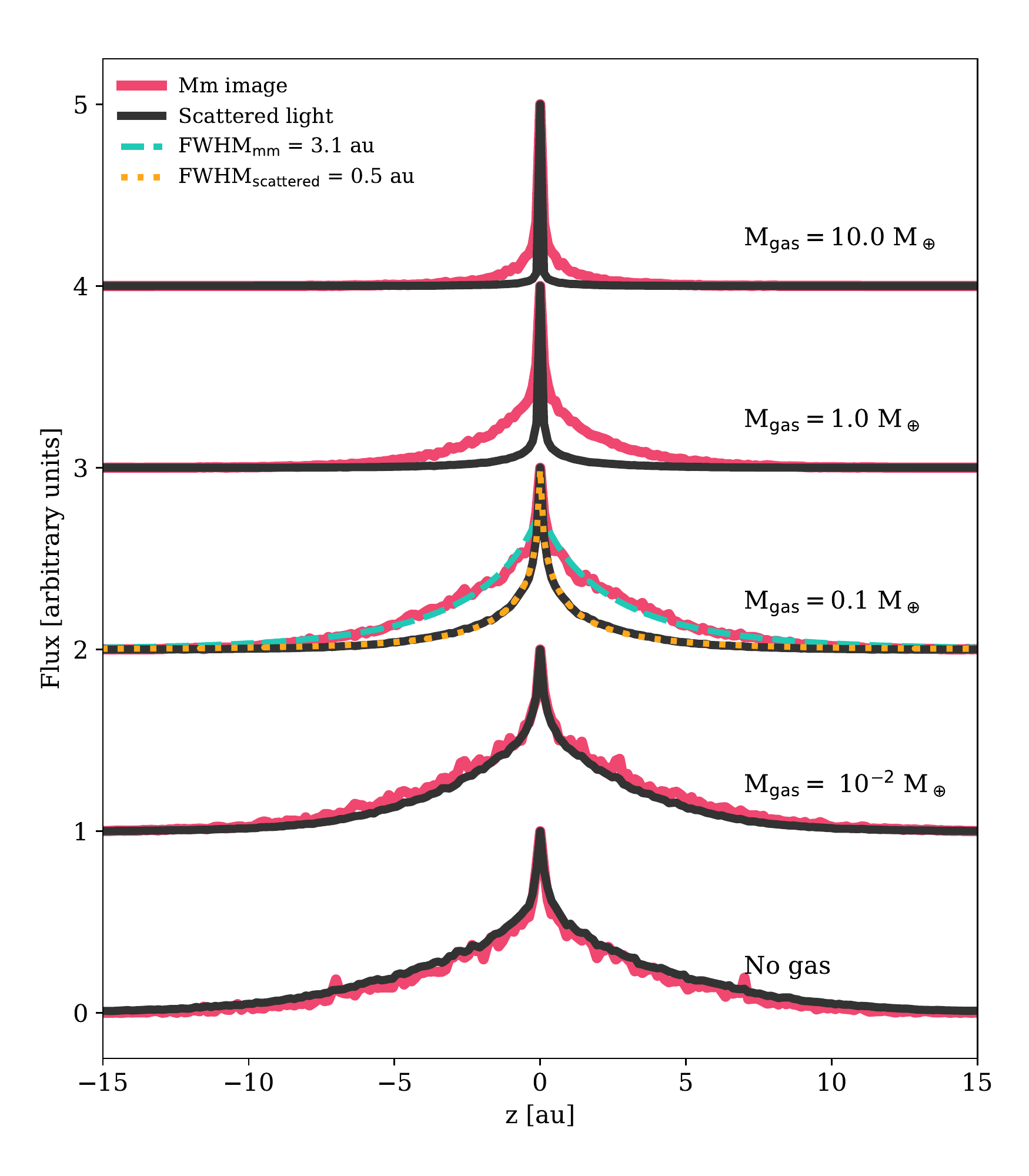}
    \caption{Normalized vertical cuts for synthetic images in scattered light and at mm wavelengths (solid dark and red lines, respectively) for simulations with different gas masses (\#1, \#4, \#5, \#11, and \#12). The vertical cut is the average over $\pm 9$ pixels around the center of the image. For the simulation with a gas mass of $0.1$\,$M_\oplus$ we fitted exponential fall-off profiles for both scattered light and mm images, and the FWHM are reported in the legend.}
    \label{fig:profiles}
\end{figure}

Figure\,\ref{fig:profiles} shows vertical cuts measured on scattered light and mm wavelength synthetic images (black and red solid lines respectively), for simulations with different amount of gas (\#1, \#4, \#5, \#11, and \#12). The vertical cuts are averaged over $9$ pixels on each side ($\sim 2.3$\,au in width) of the central pixel of the images (projected $r = 0$). The mm profile appears slightly noisier because there are fewer large particles in the simulation compared to small particles.

Overall, the profiles all have very strong peaks in the midplane of the disk ($z = 0$), and this is not only due to the presence of gas, as we see similar shapes for the simulation \#1 without gas. We also obtain similar shapes for vertical cuts along the birth ring, at projected $r = 80$\,au. This may seem counter-intuitive as all the particles are initially released with inclinations drawn from a normal distribution with a fixed standard deviation. But the shape of the profiles can be explained by projection effects. The way we initialize the inclinations of the particles means that the disk has a constant aspect ratio $h/r$. When viewed edge-on, we are thus probing a wide range of radial distances $r$ in the line of sight, from the inner edge of the birth ring $a_\mathrm{min}$ to the outermost regions (and through both the back and front sides of the disk). Along the line of sight, we are therefore not ``seeing'' one normal distribution, but the integral of several normal distributions with standard deviations increasing proportionally to $r$ (and different weights, which depend on $r$, e.g., illumination factor, surface density distribution). Furthermore, the particles should cross the midplane twice per orbit, while they reach the maximum (or minimum) altitude once per orbit, de facto increasing their contribution in the midplane. As a consequence, the combination of the different effects results in the broadening of the vertical profile, with wings wider than a normal distribution.

Figure\,\ref{fig:profiles} also shows two best-fit solutions to the vertical scattered light and mm profiles for simulation \#5, when assuming an exponential fall-off similar to the one of Eqn.\,\ref{eqn:falloff}. Since the disk is viewed edge-on, $z$ is now integrated along the line the sight, over a wide range of $r$. Therefore, we replaced $\mathrm{tan}(\psi)r$ by a ``typical'' height $H_0$, and there are two free parameters, $H_0$ and $\gamma$, determined using the \texttt{lmfit} package (\citealp{lmfit}). The scaling of the profile is obtained by minimizing the differences between the exponential fall-off and the profile to be fitted. The profiles from the simulations are not convolved by a PSF or a beam, to preserve the resolution of $1$\,mas, and both solutions match the profiles very well. We then compute the FWHM for both solutions as $2 H_0 [-\mathrm{ln}(0.5)]^{1/\gamma}$. For the mm and scattered light images, we obtain FWHM of $3.1$ and $0.5$\,au, respectively. 

Overall, the intrinsic shape of the vertical profiles observed edge-on seems to be best reproduced using exponential profiles. When using a normal distribution to model vertical cuts, either the wings of the profile, or its peak value may not be reproduced properly. This could lead to a possible under- or over-estimation of the distribution of inclinations, which can be used to infer the dynamical properties of dust particles. However, the shape of the profile is obviously altered by either the PSF or beam of the instrument used. For instance, \citet{Engler2017} used Moffat functions to model vertical cuts of SPHERE/ZIMPOL observations of the disk around HIP\,79977, such profiles having broader wings than normal distributions.

\subsubsection{Scale height in scattered light images}

\begin{figure*}
	\includegraphics[width=\hsize]{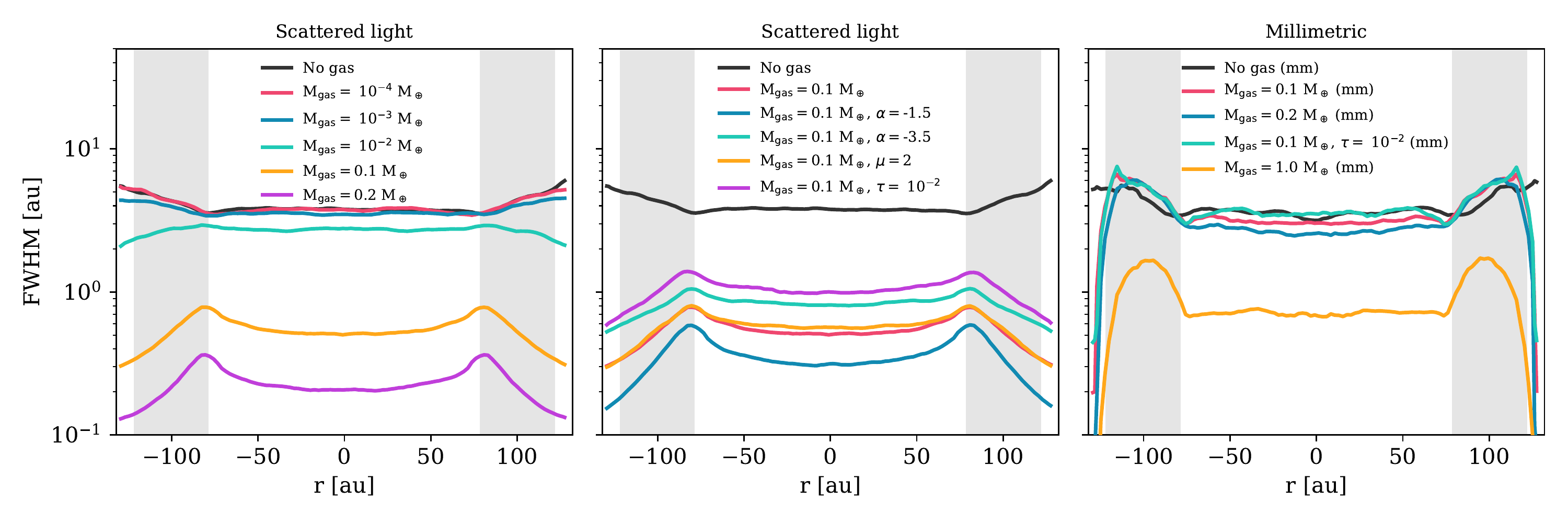}
    \caption{FWHM for vertical cuts as a function of the separation, for different simulations (scattered light for the left and center panels, and millimetric for the right panel). The $y$-axis is the same for all panels and is in log. The shaded gray areas show the location of the birth ring between $a_\mathrm{min}$ and $a_\mathrm{max}$.}
    \label{fig:disk_height}
\end{figure*}

Motivated by the previous analysis, we fitted exponential fall-off profiles to vertical cuts at different radial separations. We used the edge-on synthetic images, for all the simulations, both in scattered light and mm thermal emission, and the vertical cuts are averaged over $18$ neighboring pixels ($2.3$\,au). The left panel of Figure\,\ref{fig:disk_height} shows the radial profile of the FWHM, in units of au, for varying gas masses. The $1 \sigma$ uncertainties on the FWHM are plotted as shaded regions, but are overall smaller than the line width. The shaded gray areas show the location of the birth ring in the simulations. We see that the larger the amount of gas, the more efficiently the small particles settle towards the midplane of the disk, but this effect only starts to become noticeable for gas masses larger than $\sim 10^{-2}$\,$M_\oplus$. 

We see that, in scattered light, the profile in the regions beyond $78.5$\,au (the inner edge of the birth ring, $a_\mathrm{min}$) turns from an outwardly increasing FWHM to an outwardly decreasing one as the mass of gas increases. This is because, for projected separations larger than the inner edge of the birth ring, the line of sight intercepts deeper columns in the disk compared to separations smaller than $a_\mathrm{min}$. Therefore, in simulation \#1 (no gas), the increase in FWHM only reflects the constant aspect ratio of the disk ($\tan \psi = h/r$). When the gas mass is different from zero, the vertical settling starts to become more efficient (e.g., cyan line on the left panel of Fig.\,\ref{fig:disk_height}) and as a consequence the FWHM of the vertical cuts also decreases at larger separations and this effect is more pronounced as the gas mass increases.

The middle panel of Figure\,\ref{fig:disk_height} explores the influence of the slope $\alpha_\mathrm{gas}$ of the gas density distribution, the impact of the mean molecular weight $\mu$, and of the normalization factor for the optical depth $\tau$. Except for simulation \#1, in which there is no gas, the gas mass for all the other simulations is $0.1$\,$M_\oplus$. We find that the value of $\mu$ does not have a strong impact on the vertical structure of the disk. However, varying $\alpha_\mathrm{gas}$ or changing $\tau$ from $5 \times 10^{-3}$ to $10^{-2}$ does change the FWHM for the vertical scale height of the disk (even though we mentioned that the gas-to-dust ratio was not a critical parameter, the total dust mass does influence the collisional lifetime of the particles and therefore the final scale height of the disk). Therefore, measuring the vertical scale height of debris disks from scattered light observations cannot easily be used as an indirect way to infer the amount of gas in the system. This is even more so the case because we do not know the initial distribution of inclinations of the dust particles (fixed to $0.05$\,radians for all simulations).

As a final remark, as shown in Figure\,\ref{fig:edgeon}, because of the more efficient vertical settling of the particles, the midplane density \textit{beyond the birth ring} is larger when gas is considered, making the midplane brighter. This explains the ``needle''-like structure at large separations as the midplane is denser and hence brighter than in the simulation without gas which have a constant opening angle (hence diluting the flux, see also Fig.\,\ref{fig:zcut}). 

\subsubsection{The difference between scattered light and mm images}

We performed the same exercise on synthetic mm images, and the results are shown in the rightmost panel of Fig.\,\ref{fig:disk_height} for simulations \#1, \#5, \#6, \#10, and \#11 to test the influence of both the gas mass and $\tau$. Even though the profiles are slightly noisier, all the profiles (except the last one for $M_\mathrm{gas} = 1$\,$M_\oplus$, see later) look quite similar with each other. For simulations \#1, \#5, \#6, and \#10, this means that the presence of gas does not affect the appearance of the disks as observed at mm wavelengths. 

\begin{figure}
	\includegraphics[width=\hsize]{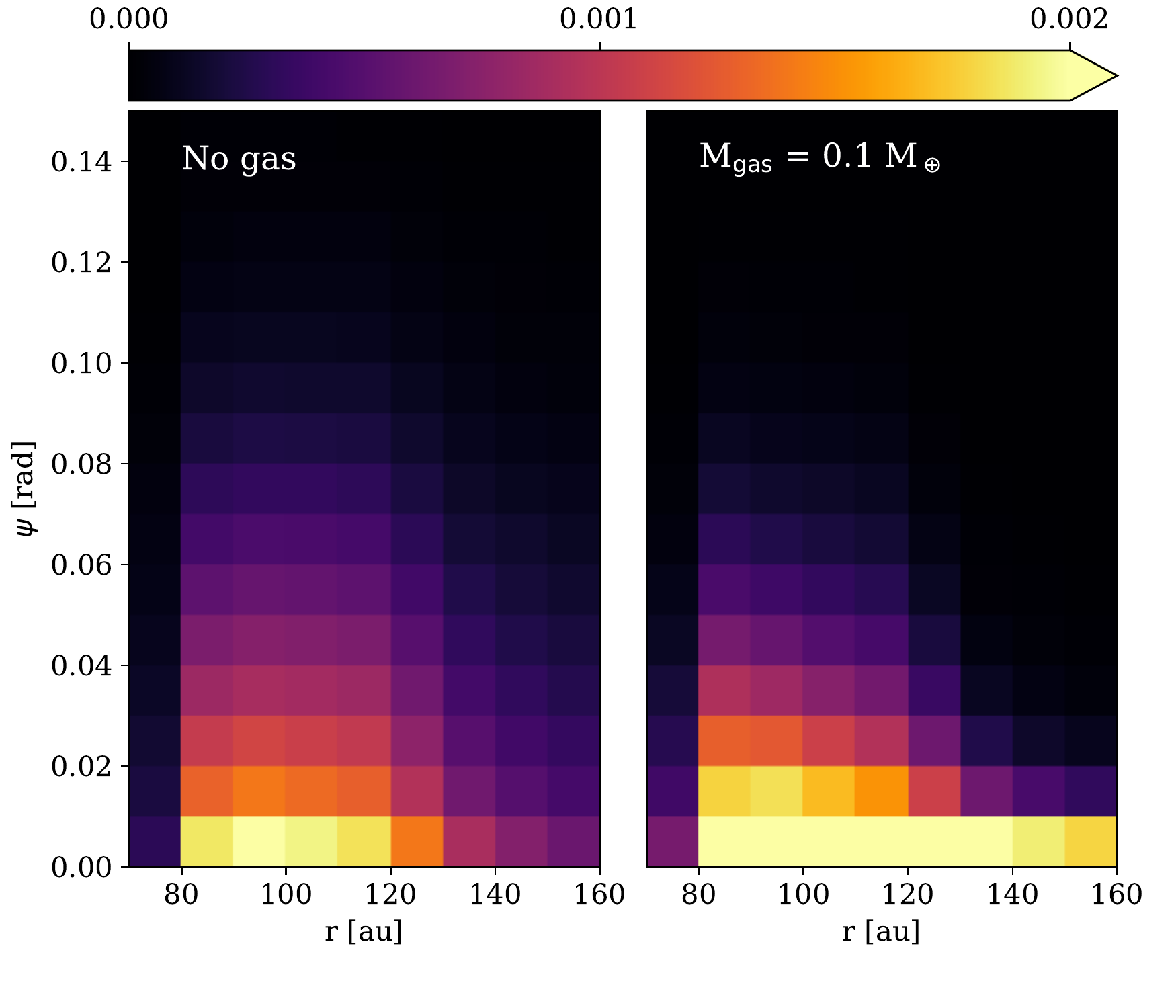}
    \caption{Zoom on the optical depth maps near the birth ring for simulations with no gas (left panel) and with a gas mass of $0.1$\,$M_\oplus$ (right panel). The scaling is linear and the same for both panels.}
    \label{fig:tau_compare}
\end{figure}

To better understand this result, in Figure\,\ref{fig:tau_compare} we show the optical depth maps of simulations \#1 and \#5 (no gas and $0.1$\,$M_\oplus$) in the vicinity of the birth ring, as a function of the angle $\psi= |\mathrm{arctan}(z/r)|$. While the optical depth is integrated over the whole size distribution, it is the quantity that drives the collisional lifetime of particles of all sizes in the simulation. Both panels have the same linear scaling, and show that in the range $80 \leq r \leq 120$\,au and $\psi \lesssim 0.04$\,radians the optical depth is larger than $10^{-3}$. Therefore, since the large grains will mostly be confined in or near the birth ring (Fig.\,\ref{fig:zcut}), they may be destroyed at high altitudes, and the amount of gas does not seem to make a significant difference. On the other hand, the small dust particles will quickly migrate outward, where the optical depth will be smaller, especially above the midplane. These particles will therefore be able to settle toward the midplane more efficiently before eventually being removed from the simulation, explaining why the scale height of scattered light images can be as small as what is seen in the left and middle panel of Fig.\,\ref{fig:disk_height} and as illustrated in Fig.\,\ref{fig:edgeon} .

Finally, for the change in FWHM to become noticeable in mm images, the gas mass needs to be larger than a certain threshold, that should be between $0.1$ and $1$\,$M_\oplus$. As a matter of fact, on Fig.\,\ref{fig:disk_height} we could not compute the FWHM for simulation \#12 because the disk becomes too flat. As shown on the top curves of Fig.\,\ref{fig:profiles}, both profiles are extremely narrow for $M_\mathrm{gas} = 10$\,$M_\oplus$, and attempts at fitting exponential profiles failed\footnote{Decreasing the pixel scale of the images from $1$\,mas to $0.5$\,mas did not change the problem.}; most of the signal being contained in one pixel of the vertical cuts. With their larger gas masses and smaller value of $\mu$, simulations \#11 and \#12 could be more representative of systems with primordial gas. This could be an interesting diagnostic to better constrain the origin (and total mass) of the gas; if the disk appears very thin at mm wavelengths, our results would suggest the gas is of primordial origin. For systems where the gas is thought to be of secondary origin, the total gas mass is expected to be smaller, and the vertical scale height of the disk should not be affected by the gas drag at mm wavelengths.

\subsection{Scattered light surface brightness profiles}

\begin{figure*}
	\includegraphics[width=\columnwidth]{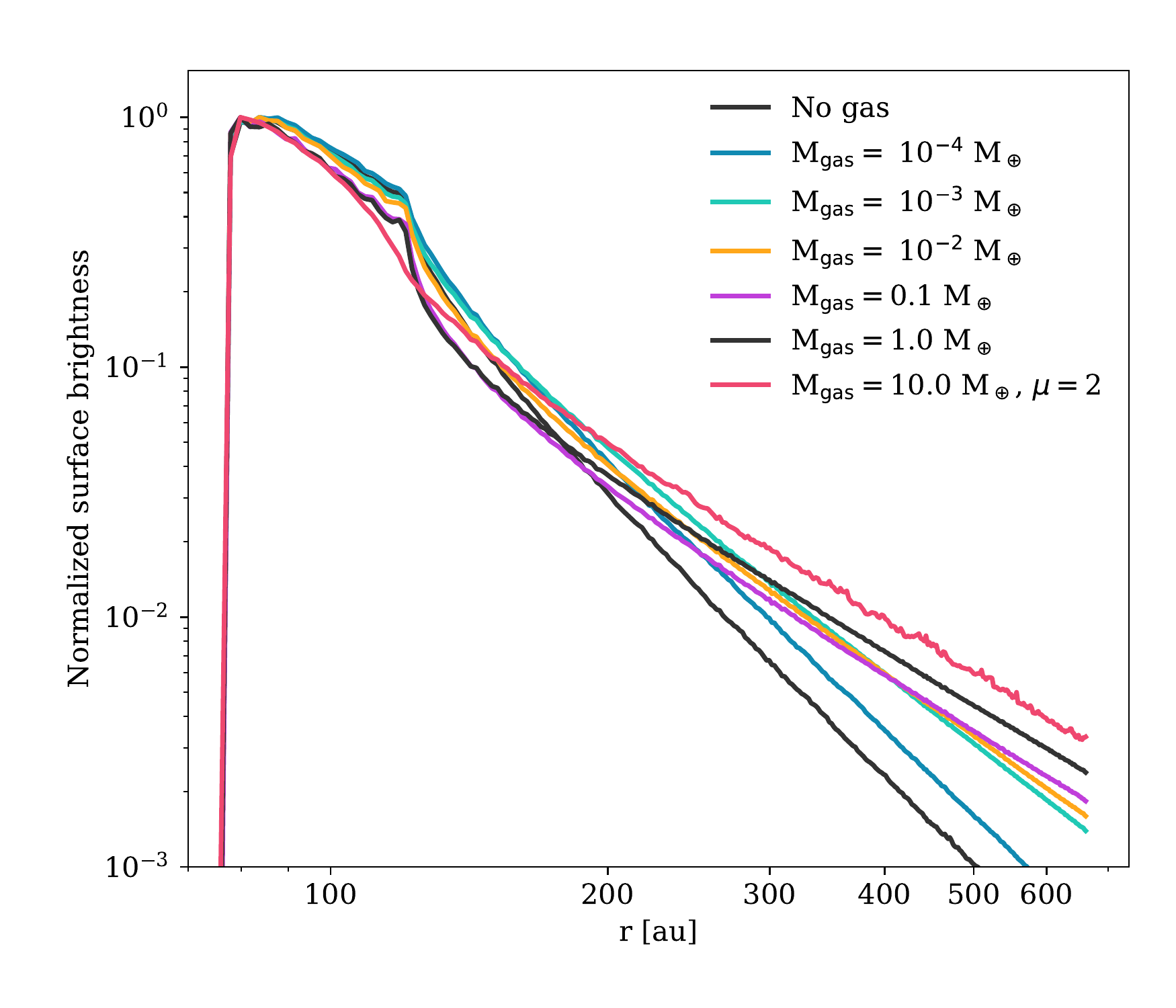}
	\includegraphics[width=\columnwidth]{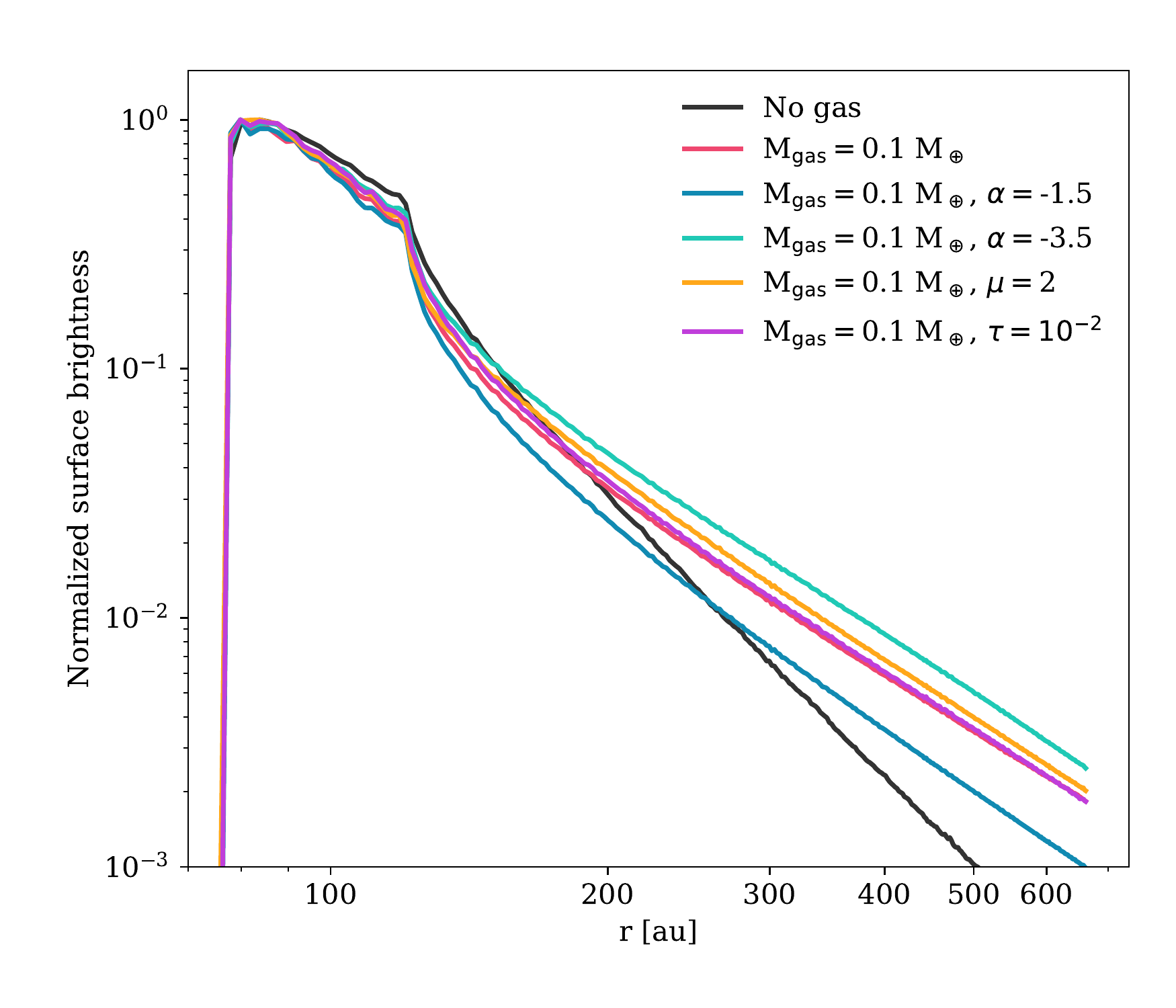}
    \caption{Normalized scattered light surface brightness distributions for different simulations. Left panel shows the distributions for a constant $\alpha_\mathrm{gas} = 2.5$, for different masses of gas, while the right panel shows the distributions for different values of $\alpha_\mathrm{gas}$, $\mu$, and $\tau$. 
    }
    \label{fig:sb}
\end{figure*}

The left panel of Figure\,\ref{fig:sb} shows that, for simulations with little or no gas (\#1 and \#2), the surface brightness (SB) follows a radial profile that tends towards $r^{-3.5}$ beyond the main belt. This is the expected behavior for a collision-dominated disk of particles only affected by stellar radiation-pressure (\citealp{Strubbe2006}, \citealp{Thebault2008}) and this results validates our approach.

However, as the gas mass increases, the radial profile becomes shallower and appears to converge towards a $r^{-2.25}$ slope in the outer regions for all $M_\mathrm{gas} > 0.1$\,$M_\oplus$ cases. This is an expected result because, contrary to the gas-free case for which all particles with non-zero $\beta$ are set on fixed eccentric orbits that make them repeatedly come back to the high-density regions of the birth ring where they eventually will get collisionally destroyed, gas drag steadily increases the pericenter of most non-zero $\beta$ particles (Fig.\,\ref{fig:three_sizes}), which thus will never come back to to the birth ring. As a matter of fact, the $\mu=28$ curve of Fig.\,\ref{fig:stability} shows that all grains released from the birth ring with $\beta > 0.005$ migrate outward until they reach a radial distance where $\beta = \eta$. As a result small grains, which dominate the flux in scattered light, will tend to populate more distant regions where they will survive longer, hence the shallower slope of the surface brightness radial profile for higher gas masses.

The fact that the radial slope of the SB profile then remains relatively constant for gas masses higher than $\simeq 0.1$\,$M_\oplus$ is also expected. Indeed, for a given value of $\mu$, the radial location of the $\beta = \eta$ equilibrium does not depend on gas masses. The only thing that changes with $M_\mathrm{gas}$ is the \textit{speed} at which this location is reached, and this speed is only governed by the total gas mass, not the gas-to-dust ratio. For low $M_\mathrm{gas}$ values, outward drag is slow and most particles get collisionally destroyed before they reach the $\beta = \eta$ point, whereas more and more grains are able to reach this location for increasing gas masses (see Fig.\,\ref{fig:sb_binned}, for which particles with $\beta$ larger than $0.002$ should migrate outward). However, once $M_\mathrm{gas}$ is high enough to allow most high-$\beta$ grains to reach their ``final destination'' at $\beta = \eta$ before collisional destruction, further increasing the gas mass will not result in major changes of the dust distribution, hence the observed asymptotic behavior in $r^{-2.25}$ that we observe for all $M_\mathrm{gas} > 0.1$\,$M_\oplus$ cases.

The profiles that we derive in this paper slightly differ from the ones reported in \citet{Krivov2009} who found similar slopes for the brightness profiles whether there is gas in the disk or not. Those differences could arise from two different assumptions. First, \citet{Krivov2009} assumed the gas density to be constant in the vertical direction (with a semi-opening angle of $16=5^{\circ}$), while we assume a normal distribution in the vertical direction. Second, they set $\mu = 2$ while in most cases we are using $\mu = 28$. This choice has an impact on two fronts; it impacts the vertical distribution of the gas (Eqn.\,\ref{eqn:hgas}), resulting in smaller $h_\mathrm{gas}$ for larger $\mu$ values, and it also impacts the maximum grain size that will migrate outwards. As shown in Fig.\,\ref{fig:stability}, larger particles will migrate outwards for larger $\mu$ values. The fact that more particles are expected to migrate outwards (because of the value of $\mu$), and that they can more efficiently do so (due to a larger density in the midplane of the disk) should explain the differences for the surface brightness profiles between this work and \citet{Krivov2009}.

Finally, still on the topic of radial profiles, it should be noted that for the mm images, all the flux is contained within the birth ring (see Fig.\,\ref{fig:sb_mm} in App.\,\ref{sec:app_sb}), even for simulation \#12 with the largest gas mass. Therefore, outward migration caused by gas drag cannot solely explain the halo of mm emission reported for HD\,32297 by \citet{MacGregor2018}.

\section{Discussion}\label{sec:discussion}

\subsection{Near-IR constraints on the vertical scale height}\label{sec:irscale}

\begin{figure}
	\includegraphics[width=\columnwidth]{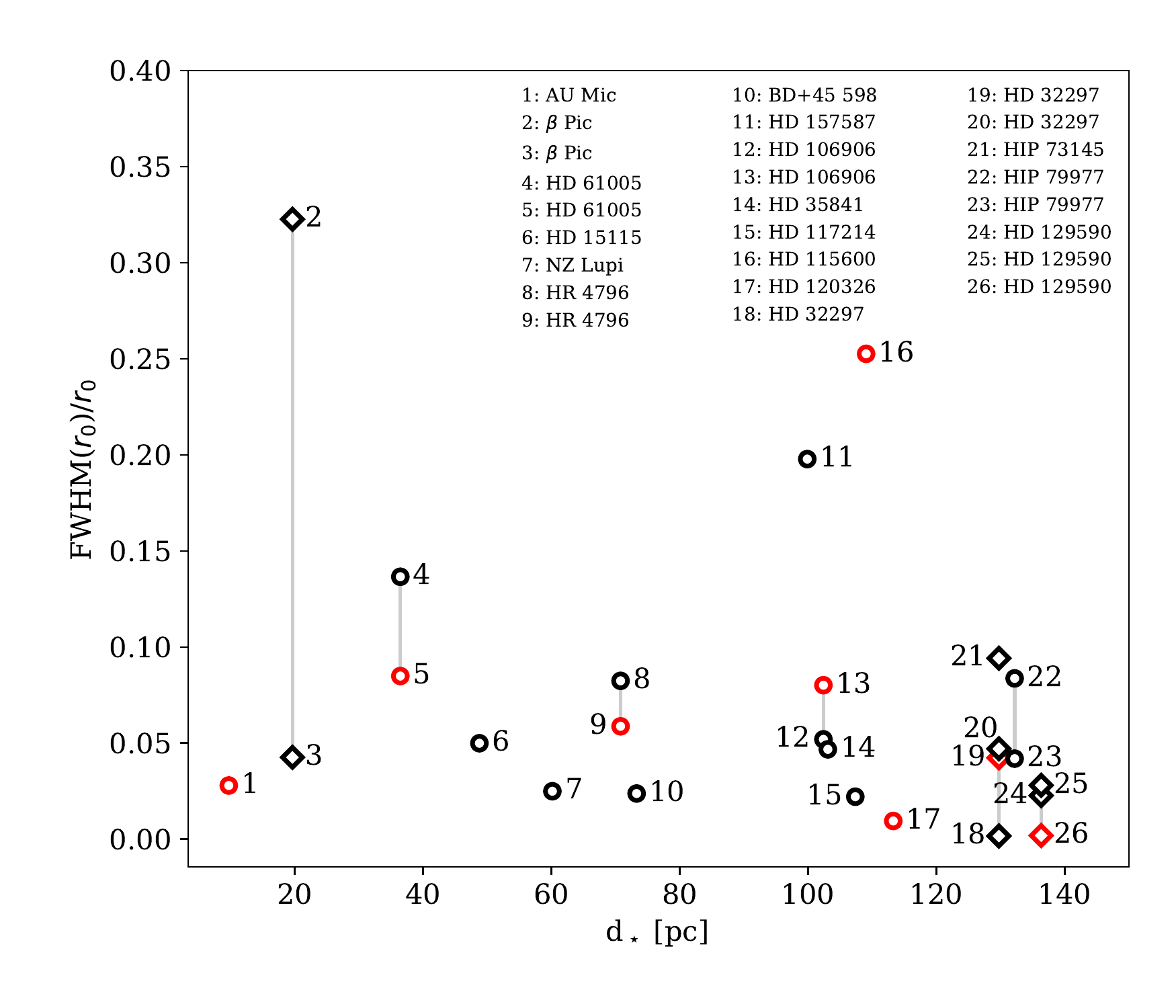}
    \caption{Aspect ratio (FWHM($r_0$)/$r_0$) of debris disks as a function of their distances $d_\star$, from the literature and this work. Stars that have several entries are connected by a thin vertical line. Values determined in this work are shown in red. Disks that are known to harbor gas are shown with a diamond symbol.}
    \label{fig:bib}
\end{figure}

\begin{table*}
	\centering
    \caption{Best-fit parameters for the vertical density distribution of debris disks, from the literature and this work.}
	\label{tab:scattered}
	\begin{tabular}{lccccccccc}
		\hline
        \# & Name        & $d_\star$ & $h_0$ & $r_0$& $\gamma^{a}$ & $\beta$ & Profile type & Gas detection  & Reference \\
           &             & [pc]      & [au]  & [au] &          &         &        & & \\
		\hline
1 & AU\,Mic & 9.714 $\pm$ 0.002 & 2.1 & 100 & 0.9 & 1 & Exponential &    & This work\\
2 & $\beta$\,Pic & 19.635 $\pm$ 0.057 & 13.7 & 100 & 2 & 1 & Gaussian & Y & \citet{Millar-Blanchar2015}\\
3 & $\beta$\,Pic & 19.635 $\pm$ 0.057 & 5.1 & 115 & 0.5 & 1.5 & Exponential & Y & \citet{Milli2014}\\
4 & HD\,61005 & 36.451 $\pm$ 0.021 & 5.8 & 100 & 2 & 1 & Gaussian &    & \citet{Olofsson2016}\\
5 & HD\,61005 & 36.451 $\pm$ 0.021 & 5.1 & 100 & 2 & 1 & Exponential &    & This work\\
6 & HD\,15115 & 48.774 $\pm$ 0.070 & 3 & 100 & 2 & 1 & Exponential &    & \citet{Engler2019b}\\
7 & NZ\,Lupi & 60.143 $\pm$ 0.080 & 1.5 & 100 & 2 & 1 & Exponential &    & \citet{Boccaletti2019}\\
8 & HR\,4796 & 70.771 $\pm$ 0.236 & 3.5 & 100 & 2 & 1 & Gaussian &    & \citet{Olofsson2020}\\
9 & HR\,4796 & 70.771 $\pm$ 0.236 & 4.1 & 100 & 1.1 & 1 & Exponential &    & This work\\
10 & BD+45\,598 & 73.271 $\pm$ 0.123 & 1 & 70 & 2 & 0 & Exponential &    & \citet{Hinkley2021}\\
11 & HD\,157587 & 99.867 $\pm$ 0.225 & 8.4 & 100 & 2 & 1 & Gaussian &    & \citet{Millar-Blanchaer2016}\\
12 & HD\,106906 & 102.382 $\pm$ 0.194 & 4.3 & 100 & 0.715 & 1 & Exponential &    & \citet{Crotts2021}\\
13 & HD\,106906 & 102.382 $\pm$ 0.194 & 4.7 & 100 & 2.3 & 1 & Exponential &    & This work\\
14 & HD\,35841 & 103.075 $\pm$ 0.139 & 2.7 & 60 & 0.560 & 1 & Exponential &    & \citet{Esposito2018}\\
15 & HD\,117214 & 107.353 $\pm$ 0.251 & 0.5 & 45.2 & 2 & 0.3 & Lorenztian &    & \citet{Engler2019}\\
16 & HD\,115600 & 109.040 $\pm$ 0.250 & 13.2 & 100 & 8.3 & 1 & Exponential &    & This work\\
17 & HD\,120326 & 113.270 $\pm$ 0.378 & 0.5 & 100 & 7.3 & 1 & Exponential &    & This work\\
18 & HD\,32297 & 129.734 $\pm$ 0.545 & 0.1 & 100 & 2 & 1 & Exponential & Y & \citet{Duchene2020}\\
19 & HD\,32297 & 129.734 $\pm$ 0.545 & 2.2 & 100 & 9.6 & 1 & Exponential & Y & This work\\
20 & HD\,32297 & 129.734 $\pm$ 0.545 & 2 & 100 & 2 & 1 & Gaussian & Y & \citet{Bhowmik2019}\\
21 & HIP\,73145 & 129.739 $\pm$ 0.468 & 4 & 100 & 2 & 1 & Gaussian & Y & \citet{Feldt2017}\\
22 & HIP\,79977 & 132.186 $\pm$ 0.414 & 3.2 & 53 & 1 & 1 & Exponential & ? & \citet{Goebel2018}\\
23 & HIP\,79977 & 132.186 $\pm$ 0.414 & 2.3 & 73 & 0.9 & 2.2 & Exponential & ? & \citet{Engler2017}\\
24 & HD\,129590 & 136.322 $\pm$ 0.440 & 1 & 73.3 & 2 & 0 & Exponential & Y & \citet{Matthews2017}\\
25 & HD\,129590 & 136.322 $\pm$ 0.440 & 1 & 59.3 & 2 & 0 & Exponential & Y & \citet{Matthews2017}\\
26 & HD\,129590 & 136.322 $\pm$ 0.440 & 0.1 & 100 & 9.7 & 1 & Exponential & Y & This work\\
      \hline
      \multicolumn{10}{l}{$^a$ In many cases, $\gamma$ is fixed and not a free parameter of the modeling. }\\
	\end{tabular}
\end{table*}

In Section\,\ref{sec:obs} we presented modeling results for eight highly inclined disks. To avoid possible biases, we only selected disks that have been observed with the SPHERE instrument with the dual-beam polarimetric imaging mode. Out of the eight systems, only two have been reported to harbor gas (HD\,32297 and HD\,129590). Unfortunately, both of the gas-bearing disks are actually among the farthest stars in our sample, and are not spatially resolved in the vertical direction in the SPHERE observations. Several other studies have attempted to estimate the vertical structure of other debris disks, some of them known to harbor gas. Let us here collect the main results of those studies to try and compare our results with a larger but less homogeneous sample. The main drawback is that those studies used near-IR observations obtained using different instruments, different techniques (either in ADI or DPI), and that the modeling was performed using different parametrizations for the vertical structure, either gaussian, exponential fall-off, or Lorentzian profiles. 

Table\,\ref{tab:scattered} summarizes, for all systems whose vertical structure has been fitted, the best-fit values for the vertical distribution reported in the literature and in this work as well as whether the disks are known to have gas. In Table\,\ref{tab:scattered}, we did not include studies where the scale height was fixed and not considered as a free parameter. Figure\,\ref{fig:bib} shows for all 26 entries the estimated aspect ratio as a function of stellar distance. Since different studies used different parametrizations, in Fig.\,\ref{fig:bib}, the aspect ratio is computed as the FWHM at the reference radius of the disk $r_0$, divided by $r_0$\footnote{Some works included a flaring index $\beta$, $H(r) = h_0 (r/r_0)^{\beta}$, but since we compute the FWHM at $r_0$ the value of $\beta$ does not matter.} The symbols for stars that have several entries are connected by a thin gray line.

We first mention two points for which the scale height might have been over- or under-estimated, as discussed by the authors of those studies. First is point \#2 from \citet[][on observations of $\beta$\,Pictoris]{Millar-Blanchar2015} and second is point \#18, from \citet[][on HD\,32297]{Duchene2020}. For HD\,32297, \citet{Duchene2020} discussed the small aspect ratio ($h/r \sim 0.1-0.2\%$) that they derived from their modeling, and mentioned that this could either be due to the use of a Gaussian profile, or could be related to the instrumental PSF used when modeling the observations. As a matter of fact, in their residual image the spine of the disk remains visible, which we also see in our residuals (Fig.\,\ref{fig:all_first}). The use of a Gaussian profile does not seem to be a fundamental issue, since we let $\gamma$ as a free parameter and found $\gamma \sim 9.6$. One possible explanation would be that the vertical structure of the disk cannot be reproduced using one single profile, possibly suggesting the existence of a narrower component. Given that both the Gemini Planet Imager (GPI, \citealp{Perrin2015}) and the SPHERE observations do not resolve the disk in the vertical direction (Section 3.3 of \citealp{Duchene2020} and Section\,\ref{sec:obs} of this paper), wanting for observations at even higher angular resolution (e.g., optical interferometry).

Regarding the disk around $\beta$\,Pictoris, \citet{Millar-Blanchar2015} found a surprisingly large aspect ratio from the modeling of the GPI observations ($h_0 = 13.7$\,au at $r_0 = 100$\,au), and the authors mention the possible impact of polarized back scattering that could artificially increase the aspect ratio if not accounted for in the modeling. Furthermore, the complex structure of the disk in the innermost regions, with the warped inner disk, is an additional challenge to constrain the vertical structure of the disk. At larger separations (and at a different wavelength, $L^\prime$ compared to $H$ band) , \citet{Milli2014} found a scale height $h_0 \sim 5.1$\,au at $r_0 = 115$\,au (point \#3 in Fig.\,\ref{fig:bib}), much smaller than the value inferred by \citet{Millar-Blanchar2015}. 

We also note that point \#16 (HD\,115600) from this work has a rather large aspect ratio, but this seems in agreement with Fig.\,\ref{fig:all_first}, which shows that the disk appears quite thick vertically, and Figure\,\ref{fig:height_compare} suggests that the disk is indeed resolved in the vertical direction. Therefore, we do not consider point \#16 as an outlier.

Table\,\ref{tab:scattered} shows that CO gas has been detected in the disks around $\beta$\,Pictoris (\citealp{Dent2014}), HD\,32297 (\citealp{Greaves2016}), HIP\,131835 (\citealp{Moor2015}), HD\,129590 (\citealp{Kral2020}). \citet{Lieman-Sifry2016} reported a tentative detection of gas around HIP\,79977. In Figure\,\ref{fig:bib}, those disks (shown with diamond symbols) do not seem to have a much smaller aspect ratio compared to the other disks. However, the distribution of the aspect ratio has a mean value of $\mathrm{FWHM}(r_0) / r_0 \sim 0.06$ (excluding point \#2) and the sample of stars has a mean distance of $94$\,pc. Considering a typical pixel scale of $13$\,mas/pixel ($14.14$ for GPI and $12.26$ for SPHERE), and assuming a typical reference radius of $100$\,au, the mean aspect ratio of the disks is about $1.3$ elements of resolution, suggesting that most disks are not spatially resolved in the vertical direction. This, on top of low number statistics (and redundant entries), prevents any firm conclusion on whether observed gas-bearing disks are indeed thinner than gas-free systems.

\subsection{Vertical scale height at millimeter wavelengths}

In the past years, more and more debris disks have been observed at high angular resolution with ALMA, and we are starting to spatially resolve the vertical structure for a handful of them. For instance, \citet{Kennedy2018} reported that their ALMA observations of HR\,4796 most likely have sufficient resolution to constrain the vertical scale height (though the authors remain cautious as the beam size is slightly larger than the scale height). According to their modeling results, they found an aspect ratio of FWHM/$r_0 = 0.08$, a value close to the one inferred from modeling the SPHERE observations ($0.06$ in this work, $0.08$ in \citealp{Olofsson2020}). Since no cold gas has been detected in the disk around HR\,4796 (\citealp{Kral2020}), this is in line with the results of our simulations.

\citet{Daley2019} observed the disk around AU\,Mic and their modeling results suggest that the disk is marginally resolved in the vertical direction as well. They derived a FWHM of $2.9$\,au at a reference radius of $40$\,au, leading to an aspect ratio of $0.0725$. Even though the measured FWHM is about $2/3$ of the beam size, the authors compared the observations to a much thinner model, yielding a worse fit to the data, suggesting that the vertical structure of the disk can be constrained with the mm observations. Our modeling results of the SPHERE observations suggest that the disk is flatter at near-IR wavelengths (aspect ratio of $0.028$) and is resolved over several elements of resolution, though we derived a slightly higher inclination ($i \sim 89.4^{\circ}$ compared to $88.5^{\circ}$ in \citealp{Daley2019}). It is thus quite likely that the disk is thinner at near-IR wavelengths compared to sub-mm wavelengths. According to our simulations, this would be the tell-tale sign of the presence of gas in the system, but \citet{Daley2019} reported upper limits between $1.7-8.7 \times 10^{-7}$\,$M_\oplus$ for the total CO gas mass in the disk around AU\,Mic, which is much too low for gas to affect the dust distribution (see left panel of Fig.\,\ref{fig:disk_height}). Therefore, unless CO is rapidly dissociated into C and O (explaining the upper limit of CO gas), there should be another mechanism at play, acting mostly on the very small dust particles, that can make the disk thinner at near-IR wavelengths. A possible path to explore could be the (yet poorly constrained) mechanism responsible for the fast moving structures that have been detected around AU\,Mic (\citealp{Boccaletti2015,Boccaletti2018}, \citealp{Sezestre2017}).

\citet{Matra2019b} modeled ALMA observations of the disk around $\beta$\,Pictoris, putting a special emphasis on its vertical structure. They found that this vertical structure cannot be explained using just one Gaussian profile, but that the observations are best reproduced using two Gaussian profiles. They derived aspect ratio of $0.014$ and $0.11$, corresponding to FWHM$(r_0)/r_0$ of $0.033$ and $0.26$, respectively. The authors also tried to use an exponential fall-off profile, and the best fit model is obtained for $\gamma \sim 0.9$. They concluded that the exponential fall-off model could reproduce the observations as well as the model with two Gaussian profiles, if flaring was also included as a free parameter. The authors concluded that the broad wings of the vertical density distribution (either the two Gaussian or the exponential fall-off) were indicative of two populations of dust grains, with different inclinations. However, the profiles measured on our simulations (Fig.\,\ref{fig:profiles}), also show broad wings (though not convolved by a PSF or a beam), even for simulations without gas, while the inclinations of the particles are drawn from one normal distribution, suggesting that two populations might not be required to fully explain the ALMA observations.

\citet{Lovell2021} modeled ALMA observations of the disk around q$^1$\,Eri, and found a scale height $h \sim 0.05$ ($h$ being the standard deviation of a Gaussian profile), corresponding to an aspect ratio of FWHM($r_0$)$/r_0 \sim 0.12$, slightly larger than the values reported for the other disks above. \citet{Marino2016} reported an upper limit for HD\,181327, with an aspect ratio (using a Gaussian profile) that has to be smaller than $0.14$, but the inclination of the disk around HD\,181327 is not the most favorable to firmly constrain its vertical structure (though see Section\,4.9 in \citealp{Marino2016} for a discussion on the impact of the radial width, vertical extent, and inclination). 

Last but not least, the disk around HD\,32297 would be an ideal target to compare the scale height at different wavelengths. It has been observed with ALMA at different wavelengths but unfortunately has yet to be resolved in the vertical direction (\citealp{Cataldi2020}, \citealp{MacGregor2018}).

\subsection{Radial extent in scattered light observations}

Another possible diagnostic to infer the presence of gas in a debris disk would be to measure its surface brightness profile. Our simulations with gas suggest that the profile should converge to a shallower slope ($\sim -2.25$) than in simulations without gas ($-3.5$ in agreement with \citealp{Thebault2008}), though \citet{Krivov2009} found no significant differences with and without gas.

Nonetheless, this diagnostic may not be directly applicable with contemporary observations. First, in our simulations, we assumed that the gaseous disk does not have a sharp edge in the outer regions and that it extends following a power-law. The motivation for this assumption was to avoid boundary conditions effects such as the oscillations seen in the simulations of \citet{Takeuchi2001}. ALMA observations may not yet have the sensitivity to probe if the outer edge of the gas disk stops abruptly or not. \citet{Cataldi2020} reported that the gaseous disk around HD\,32297 is only mildly more spread out than the planetesimal belt, using a Gaussian profile for the gas radial density distribution. It is therefore not clear whether the gas disk can extend farther out, with a shallower density distribution, and if the ALMA observations would have had the sensitivity to detect it.

Second, even though scattered light observations of debris disks tremendously improved in the past years, thanks to new generation of instruments, they still are not sensitive enough to detect the halo of small dust grains in the outer regions. As shown in Fig.\,\ref{fig:sb}, beyond the birth ring, the surface brightness profile quickly drops of about one order of magnitude whether there is gas or not. Since the slope reaches the asymptotic behavior in $-2.5$ or $-3.5$ much farther away it remains challenging, if not impossible, to accurately measure the slope of the brightness profile. The most promising avenue remains space-based missions where the stability of the instrument, and absence of the earth's atmosphere, allow to best probe the faint outer regions (e.g., \citealp{Schneider2014,Schneider2018}). Such observations have revealed that for instance the disks around HD\,32297 and HD\,61005 display very extended wings, extending much farther away from the birth ring, but while HD\,32297 is known to host significant amount of gas, HD\,61005 does not seem to contain any detectable amount of CO gas (\citealp{Olofsson2016}).

Nonetheless, it should be noted that the disk around HD\,129590 appears quite extended in the radial direction in our SPHERE observations (Fig.\,\ref{fig:all_first}). Interestingly, the residuals show positive and negative structures in the innermost regions, close to the birth ring. This is either related to the determination of the phase function (as mentioned earlier) or a limitation of the chosen parametrization of the radial density distribution. Using two power-laws, we cannot have at the same time a narrow birth ring and an extended outward halo which seems to be the case for this disk. This target is therefore a very promising one for high angular resolution observations with ALMA, to constrain both the location of the planetesimal belt and the radial extent of the gaseous disk that was reported in \citet{Kral2020}.

\subsection{Planet induced vertical stirring}

We have here considered dust disk vertical thickness as an indirect way to constrain the amount of gas in a given debris disk, taking into account the effect of stellar gravity and radiation pressure as well as of mutual grain collisions. There are, however, additional mechanisms that could affect disk vertical structure that have been left out of our investigation. Such mechanisms could potentially interfere with our conclusions regarding the coupling between disk thickness and the amount of gas.

Perhaps the most obvious effect would be the gravitational pull of a planet or planetary embryo that could puff up the disk and counteract gas-induced vertical settling. It is beyond the scope of the present study to include such perturbing effects into our already complex numerical code, but we can make some order of magnitude estimates. We first note that our initial $\sigma_i = 0.05$\,radians (the standard deviation of the normal distribution for $i$) for grain progenitors in the parent body belt implicitly assumes that ``something'' is stirring up their orbits to this value, the most likely candidates being a string of large objects embedded in the belt with an escape speed $v_\mathrm{esc}\sim 0.05v_\mathrm{k}$ \citep{Artymowicz1997}, which would correspond to Vesta-to-Ceres-sized objects. However, it is unlikely that such embedded objects could prevent gas-induced vertical settling. For small $\mu$m-sized grains, this is because these grains are rapidly pushed out of the parent belt because of the coupling between radiation pressure and gas drag (Fig.\,\ref{fig:three_sizes}) and will thus never come in contact with the perturbing embryos. Larger, mm-sized grains will stay in the belt much longer, but the typical timescale for gravitational stirring by embedded embryos exceeds $1$\,Myr for a belt at $100$\,au \citep{Krivov2018}, whereas the timescale for vertical settling (for the $M_\mathrm{gas} > 1$\,$M_{\oplus}$ cases for which such large grains do settle) is only of the order of a few $10^{4}$ years (see Fig.\,\ref{fig:compare_incl}). Larger embedded perturbers could act faster but they would have puffed up the belt of parent planetesimals to values far exceeding the $0.05$ inclination considered here.

A planet exterior to the parent belt on an inclined orbit could also, by a combination of warping and orbital precession, vertically thicken the belt, but here again the timescales are of the order of several Myrs (\citealp{Dong2020}), highly exceeding the gas-induced settling time that we observe in our model.

\section{Summary}\label{sec:conclusion}

Contemporary observing facilities allow us to obtain spatially resolved images of debris disks with unprecedented angular resolution, both at near-IR and mm wavelengths. With the aim of constraining the vertical scale height we defined a homogeneous sample of eight disks seen close to edge-on and observed with SPHERE in polarimetric mode. We presented results of geometric modeling, and among the disks that are likely spatially resolved in the vertical direction, we find that there is a significant spread in the vertical FWHM. We also found that some of the disks seem to be very flat (albeit being the farthest from earth in our sample and spatially unresolved), and CO gas has been detected for two of them (HD\,32297 and HD\,129590) .

To quantify the possible impact of a gaseous disk on the dynamics of dust particles, we performed numerical N-body simulations that include the drag force exerted by the gaseous disk on the dust grains, as well as collisions. We found that the gas has an impact on both the radial and vertical distribution of the dust grains. For intermediate gas masses (comparable to gaseous disks of second generation,) the effect on the vertical distribution is stronger for small particles compared to larger dust grains, and therefore it is best seen in scattered light compared to mm images. For systems more akin to primordial (or hybrid) disks, with higher gas masses ($\geq 1$\,$M_\oplus$), vertical settling also becomes significant at mm wavelengths, which could provide an interesting diagnostic to better constrain the origin of the gas (provided we have observations with sufficient angular resolution). 

Under the influence of gas drag, small particles migrate outwards, where the collisional timescales are longer. Those particles can therefore survive longer and have more time to settle toward the midplane of the disk as they feel the headwind of the gas every time they cross the midplane. Nonetheless, there is a competition between outward drift and vertical settling: our results suggest that the disk is flatter for intermediate-sized particles compared to grains that are close to the blow-out size. The latter are quickly parked at large separations, where the gas density is smaller and as a consequence the vertical settling is slower. On the other hand, larger grains remain in the vicinity of the birth ring, and for intermediate gas masses ($\leq 0.1$\,$M_\oplus$), are destroyed before their inclinations can be significantly dampened by gas drag. For larger gas masses (comparable to primordial or hybrid disks), the competition between outward drift and collisional lifetime swings in the other direction, and larger grains are able to migrate outside of the birth ring before they can be destroyed, where their inclinations will decrease over time. 

We found that the intrinsic shape of the vertical profiles of our simulations are best reproduced using exponential profiles with broad wings, whether gas is present in the system or not. Regarding the radial distribution, the surface brightness profiles of gas bearing disks seem to be shallower than the profile in $r^{-3.5}$ expected when only considering gravity and radiation pressure, converging towards a profile in $r^{-2.25}$ as the gas mass increases.

Even though the two gas-bearing disks in our sample, HD\,32297 and HD\,129590, seem to be very flat, they are among the farthest stars, and the SPHERE observations do not resolve their vertical scale height. Trying to put the results of our simulations in a broader context, we gathered information from the literature, from studies trying to constrain the scale height of debris disks. Despite a larger, though less homogeneous, sample we concluded that most disks are most likely not spatially resolved vertically, preventing us from confronting our simulations with observations. 

There is therefore a need for observations at high angular resolution, both at near-IR and mm wavelengths, and the two gas-bearing disks of our sample (HD\,32297 and HD\,129590) are ideal candidates for future observations. Long-baseline ALMA observations of HD\,129590 would help better constraining the location of the birth ring and ideally its scale height. Near-IR interferometric observations of HD\,32297 may help constraining the scale height in scattered light, with higher angular resolution compared to the SPHERE observations presented here, and long baseline mm interferometry would also put more stringent constraints on the vertical structure of the disk.

\section*{Acknowledgements}
We thank the anonymous referee for a thorough review of the paper and for valuable comments that help improving this manuscript. 
Based on observations collected at the European Southern Observatory under ESO programs 598.C-0359(F), 95.C-0273(A), 97.C-0523(A), 98.C-0686(B,D), 105.20GP.011, and 101.C-0128(D).
J.\,O., A.\,B., N.\,G., and M.\,M. acknowledge support by ANID, -- Millennium Science Initiative Program -- NCN19\_171. J.\,O. acknowledges support from the Universidad de Valpara\'iso, and from Fondecyt (grant 1180395). A.\,B. gratefully acknowledges support by the ANID BASAL project FB210003. A.\,A.\,S. acknowledges financial support by the Gates Cambridge Trust (OPP1144).
This publication makes use of VOSA, developed under the Spanish Virtual Observatory project supported by the Spanish MINECO through grant AyA2017-84089. VOSA has been partially updated by using funding from the European Union's Horizon 2020 Research and Innovation Programme, under Grant Agreement nº 776403 (EXOPLANETS-A).
SPHERE is an instrument designed and built by a consortium consisting of IPAG (Grenoble, France), MPIA (Heidelberg, Germany), LAM (Marseille, France), LESIA (Paris, France), Laboratoire Lagrange (Nice, France), INAF–Osservatorio di Padova (Italy), Observatoire de Gen\`eve (Switzerland), ETH Zurich (Switzerland), NOVA (Netherlands), ONERA (France) and ASTRON (Netherlands) in collaboration with ESO. SPHERE was funded by ESO, with additional contributions from CNRS (France), MPIA (Germany), INAF (Italy), FINES (Switzerland) and NOVA (Netherlands).  SPHERE also received funding from the European Commission Sixth and Seventh Framework Programmes as part of the Optical Infrared Coordination Network for Astronomy (OPTICON) under grant number RII3-Ct-2004-001566 for FP6 (2004–2008), grant number 226604 for FP7 (2009–2012) and grant number 312430 for FP7 (2013–2016). We also acknowledge financial support from the Programme National de Plan\'etologie (PNP) and the Programme National de Physique Stellaire (PNPS) of CNRS-INSU in France. This work has also been supported by a grant from the French Labex OSUG@2020 (Investissements d'avenir – ANR10 LABX56). The project is supported by CNRS, by the Agence Nationale de la Recherche (ANR-14-CE33-0018). It has also been carried out within the frame of the National Centre for Competence in Research PlanetS supported by the Swiss National Science Foundation (SNSF). MRM, HMS, and SD are pleased to acknowledge this financial support of the SNSF. 
This work has made use of data from the European Space Agency (ESA) mission {\it Gaia} (\url{https://www.cosmos.esa.int/gaia}), processed by the {\it Gaia} Data Processing and Analysis Consortium (DPAC, \url{https://www.cosmos.esa.int/web/gaia/dpac/consortium}). Funding for the DPAC has been provided by national institutions, in particular the institutions participating in the {\it Gaia} Multilateral Agreement.
This research made use of Astropy,\footnote{\url{http://www.astropy.org}} a community-developed core Python package for Astronomy \citep{astropy:2013, astropy:2018}, Numpy (\citealp{numpy}), Matplotlib (\citealp{matplotlib}), Scipy (\citealp{scipy}), as well as Uncertainties: a Python package for calculations with uncertainties\footnote{Eric O. LEBIGOT, \url{http://pythonhosted.org/uncertainties/}}.
%

\section*{Data Availability}

The data underlying this article will be shared on reasonable request to the corresponding author. 



\bibliographystyle{mnras}




\appendix

\section{Phase function}

We here show the polarized phase functions $S_\mathrm{12}$ derived in the modeling presented in Section\,\ref{sec:obs} for all eight disks. The phase functions shown here are the average of the two phase functions estimated for each side of the disk (e.g., North and South sides). For some disks, the increases at low or high scattering angles correspond to regions close or behind the coronagraph and hence cannot be trusted.

\begin{figure*}
    \includegraphics[width=2\columnwidth]{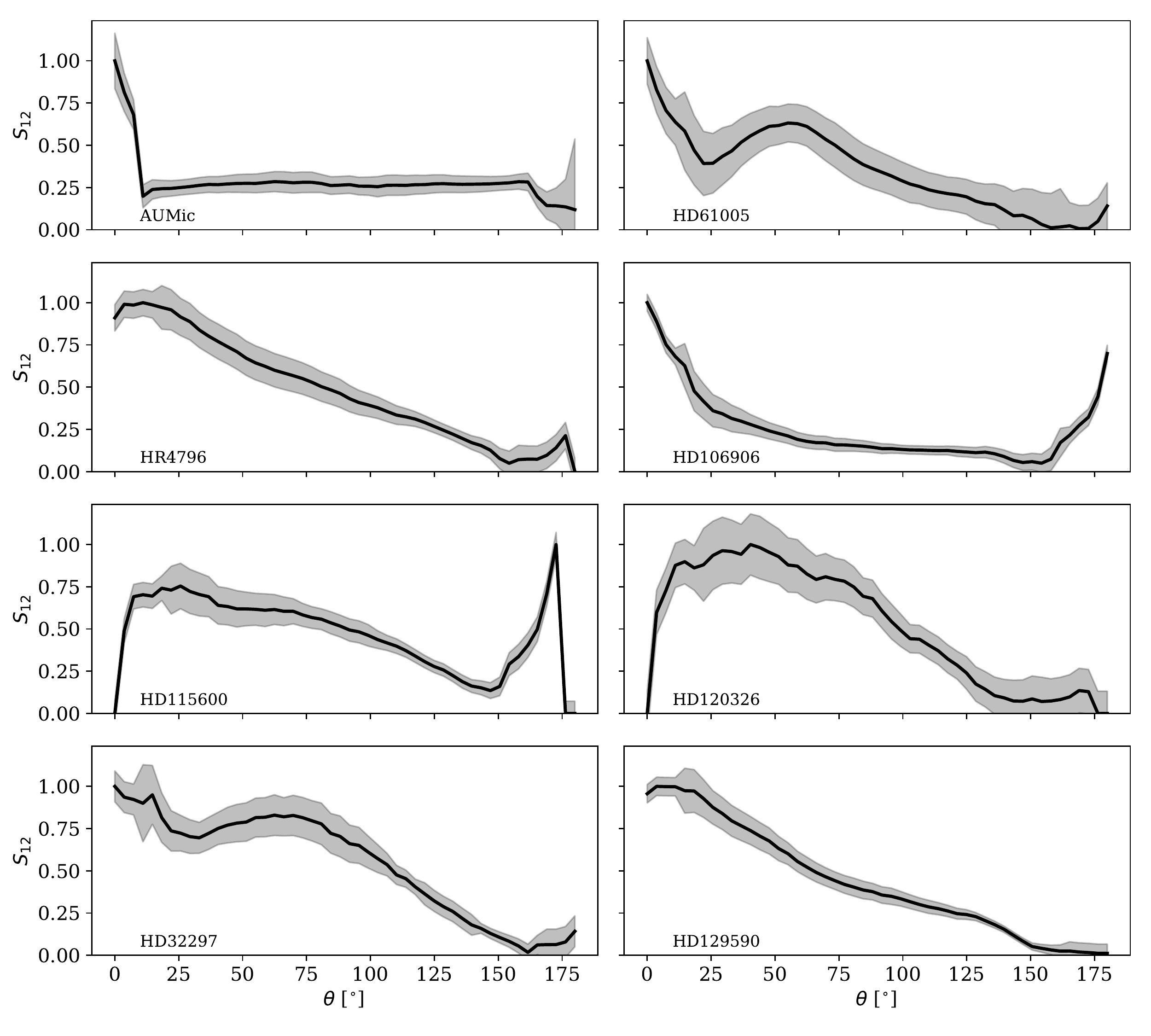}
    \caption{Polarized phase functions for all the geometric models presented in Section\,\ref{sec:obs}, normalized to unity.}
    \label{fig:pfunc}
\end{figure*}

\section{Timestep convergence}\label{sec:app_dt}

\begin{figure}
	\includegraphics[width=\hsize]{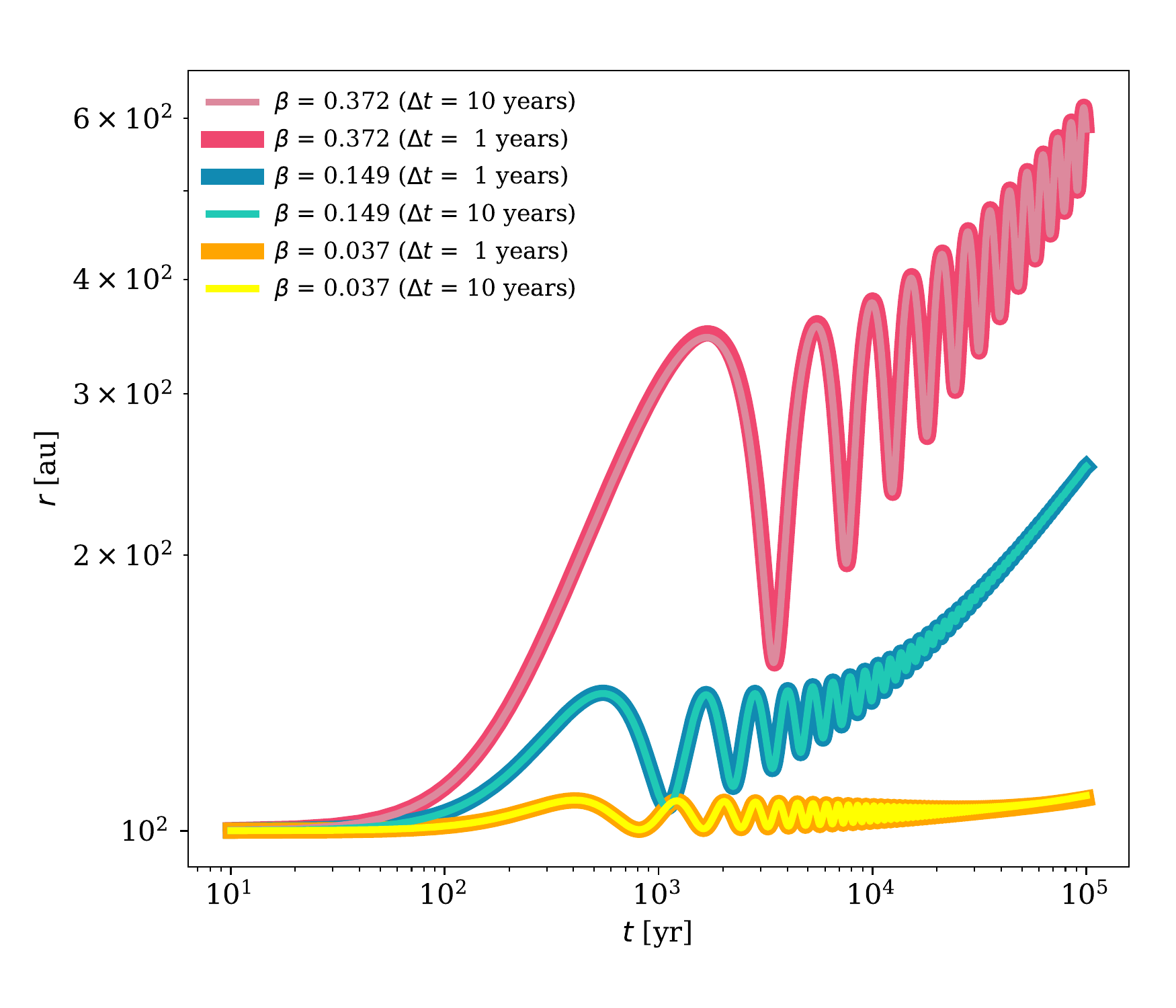}
    \caption{Distance in the midplane of particles with different $\beta$ values as a function of time, when integrating using different timesteps $\Delta t$ ($1$ and $10$\,years).}
    \label{fig:dt_check}
\end{figure}

Figure\,\ref{fig:dt_check} shows the distance in the midplane ($\sqrt{x^2+y^2}$) as a function of time, for different $\beta$ values, for different integration timesteps $\Delta t$ ($1$ and $10$\,years) for simulation \#5 ($M_\mathrm{gas} = 0.1$\,$M_\oplus$), validating our choice for the integration timestep.

\section{Inclination dampening as a function of $M_\mathrm{gas}$}

\begin{figure}
	\includegraphics[width=\hsize]{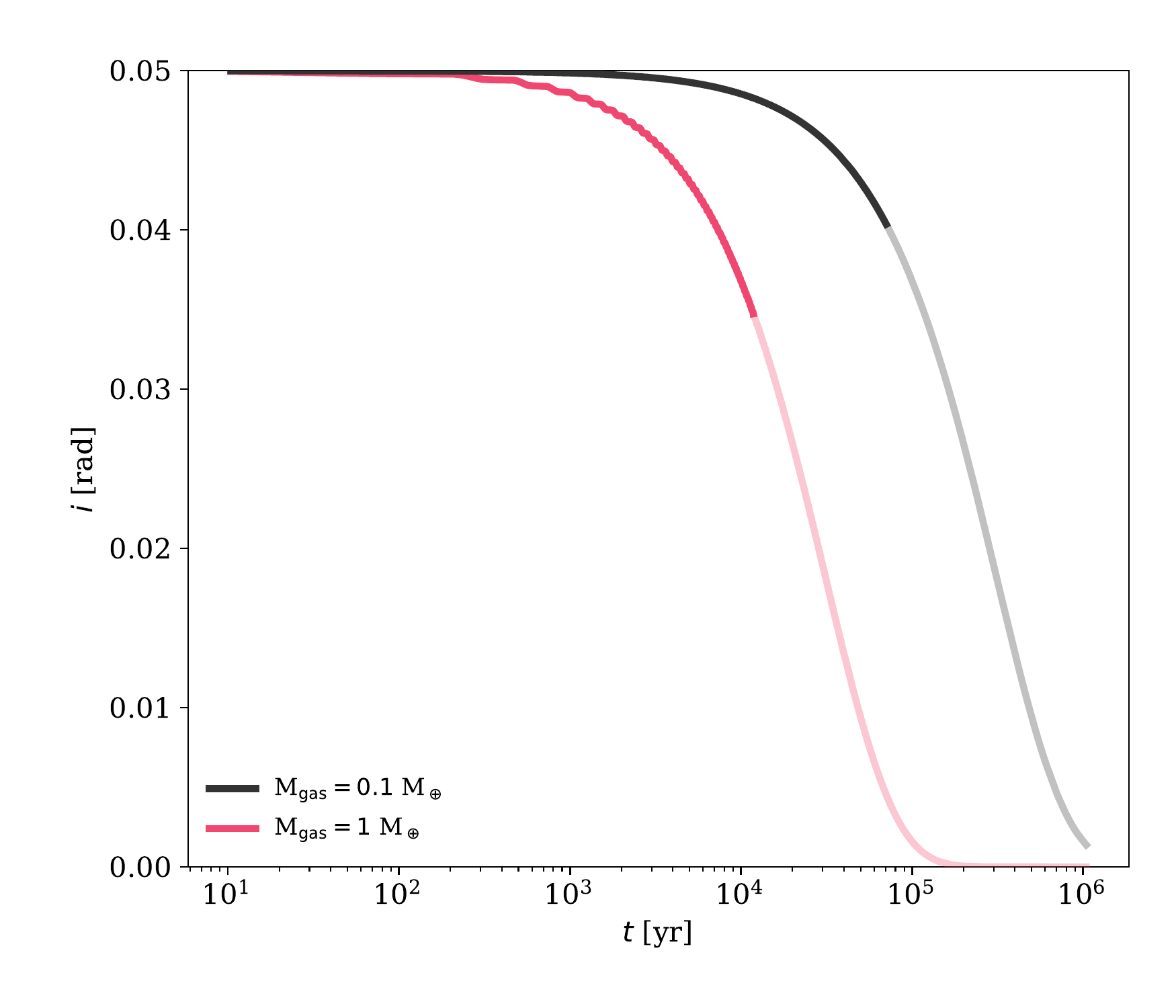}
    \caption{Inclination as a function of time for a particle with size $s = 500$\,$\mu$m for simulations \#5 and \#11 with gas masses of $0.1$ and $1$\,$M_\oplus$ (black and red, respectively).}
    \label{fig:compare_incl}
\end{figure}

Figure\,\ref{fig:compare_incl} shows the dampening of the inclination of large particles ($s = 500$\,$\mu$m) as a function of time, for simulations with two different gas masses ($0.1$ and $1$\,$M_\oplus$). The Figure shows that for the simulation with larger gas mass, the inclination of the large particle starts decreasing much earlier compared to the simulations with a smaller gas mass.

\section{Optical depth map}\label{sec:app_tau}

We show here the optical depth maps for the $19$ iterations for simulation \#5. The color scale is the same for all panels. The figure shows that the simulation reached a steady state after the $6^{\mathrm{th}}$ iteration.

\begin{figure*}
	\includegraphics[width=\hsize]{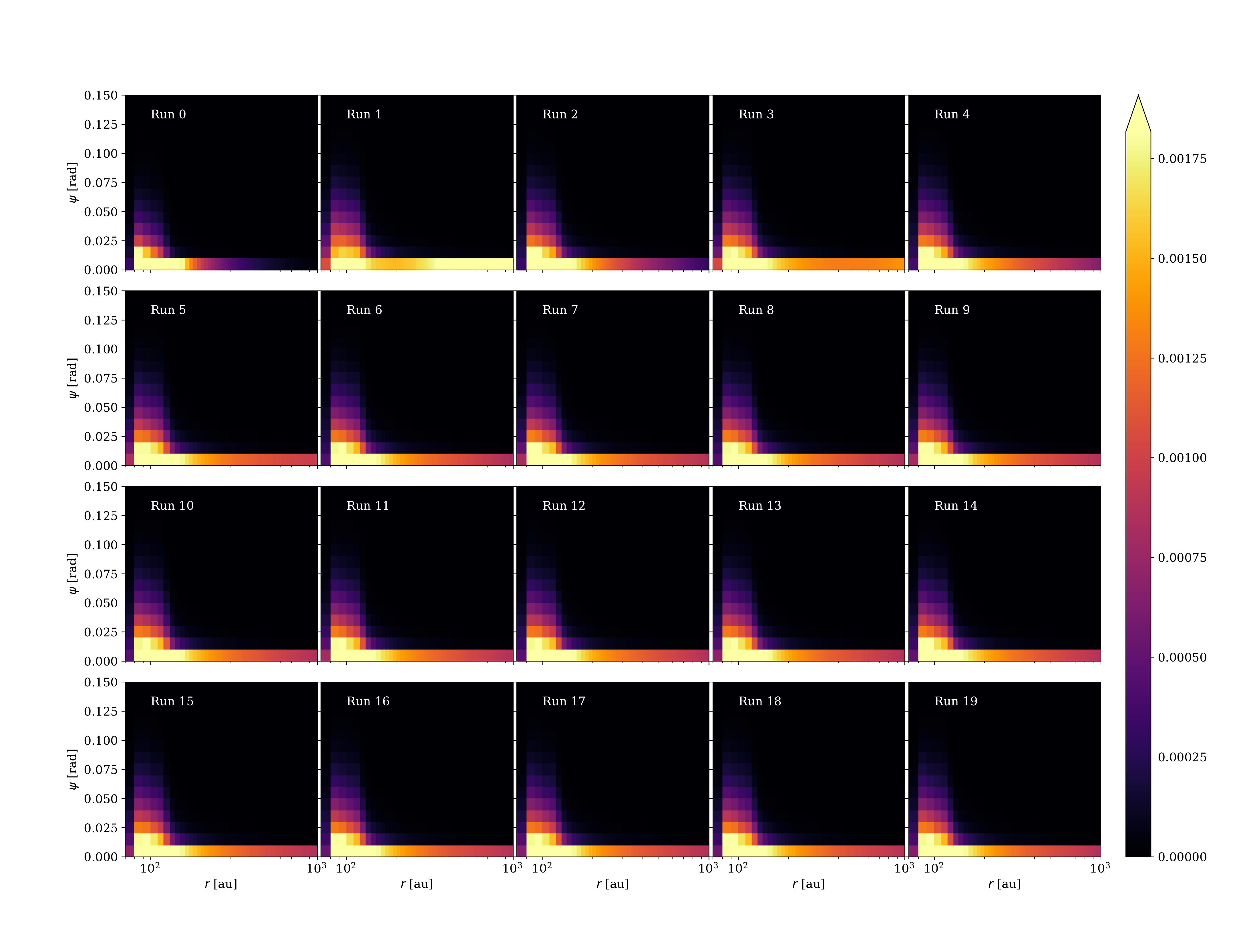}
    \caption{Optical depth map $\tau(r, \psi_\mathrm{c})$ for $19$ consecutive runs, for M$_\mathrm{gas} = 0.1$\,M$_\oplus$ and $\alpha = -2.5$ (simulation \#5). The color scale is the same for all panels, the $x$-axis is in log scale.}
    \label{fig:tau_map}
\end{figure*}

\section{Sized dependent surface brightness profiles}\label{sec:app_sb}

\begin{figure*}
	\includegraphics[width=\hsize]{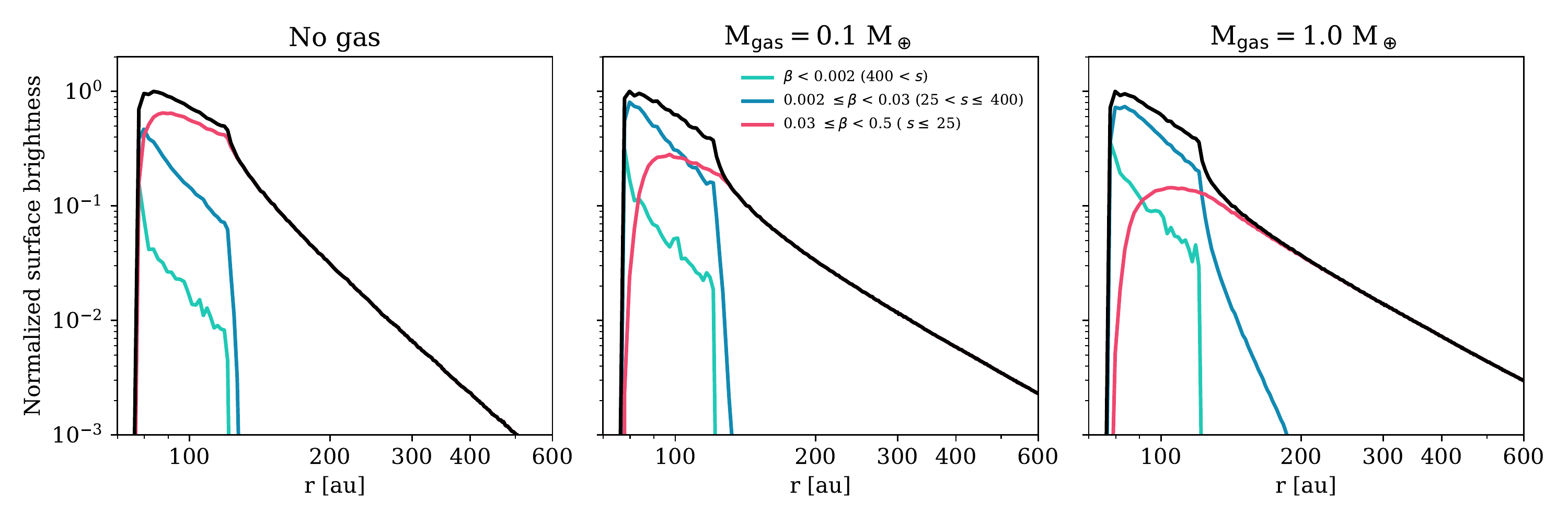}
    \caption{Normalized surface brightness profiles for simulations \#1, \#5, and \#11, showing the contributions of three different grain size intervals (see caption in the central panel).}
    \label{fig:sb_binned}
\end{figure*}

Figure\,\ref{fig:sb_binned} shows the contributions to the surface brightness profiles of three different grain size intervals: $s >400$\,$\mu$m, $25 < s \leq 400$\,$\mu$m, and $s \leq 25$\,$\mu$m. We show the results for simulations \#1, \#5, and \#11. As the gas mass increases, larger particles are migrating outwards.

\begin{figure}
	\includegraphics[width=\hsize]{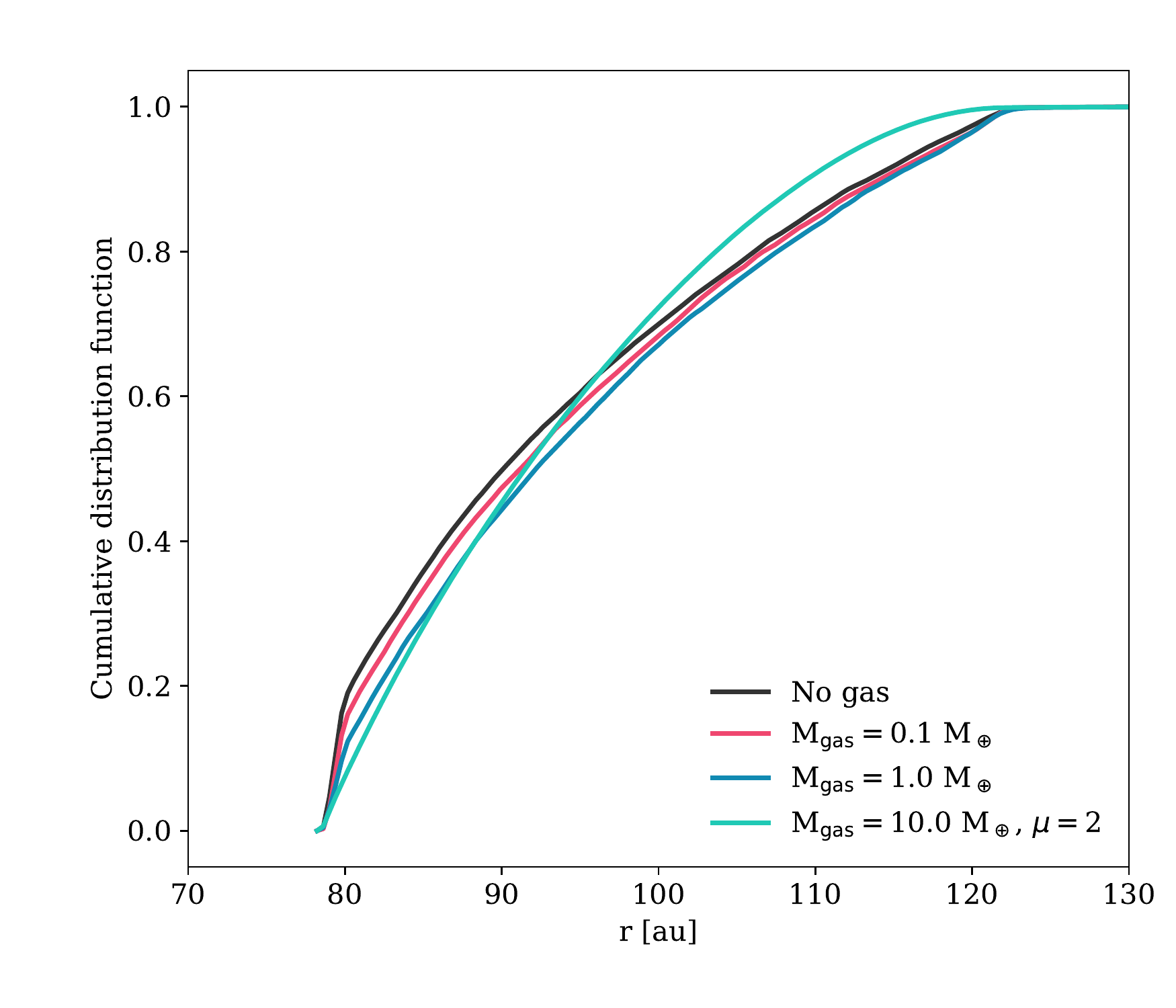}
    \caption{Cumulative distribution function of the mm flux, as a function of the stellar distance, for simulations \#1, \#5, \#11, and \#12.}
    \label{fig:sb_mm}
\end{figure}

Figure\,\ref{fig:sb_mm} shows the cumulative distribution function of the flux at mm wavelengths, as a function of the stellar distance, for simulations with and without gas, demonstrating that all the mm flux is contained within the birth ring of the disk and does not extend beyond.

\bsp	
\label{lastpage}
\end{document}